\documentclass[12pt]{article}
\usepackage{amssymb}
\def\inh{\vskip 0.075truein \noindent\hangindent=12 pt \hangafter=1}

\usepackage{amsthm}\theoremstyle{remark}

\newcommand{\bte}{\begin{quote}\begin{theorem}}
\newcommand{\ete}[1]{\label{#1}\end{theorem}\end{quote}}
\newcommand{\bcom}{\begin{quote}\end{quote}}
\newcommand{\bex}{\begin{quote}\begin{example}}
\newcommand{\eex}[1]{\label{#1}\end{example}\end{quote}}
\newcommand{\bcon}{\begin{quote}\begin{conclusion}}
\newcommand{\econ}[1]{\label{#1}\end{conclusion}\end{quote}}
\newcommand{\bdefi}{\begin{quote}\begin{definition}}
\newcommand{\edefi}[1]{\label{#1}\end{definition}\end{quote}}

\newcommand{\blem}{\begin{quote}\begin{lemma}}
\newcommand{\elem}[1]{\label{#1}\end{lemma}\end{quote}}

\newcommand{\bpr}{\begin{quote}\begin{problem}}
\newcommand{\epr}[1]{\label{#1}\end{problem}\end{quote}}

\newcommand{\f}{\frac}
\newcommand{\p}{\partial}
\newcommand{\n}{\nonumber \\}
\newcommand{\inti}{\int_{-\infty}^\infty}
\newcommand{\beq}{\begin{eqnarray}}
\newcommand{\eeq}[1]{\label{#1}\end{eqnarray}}
\newcommand\eq[1]{(\ref{#1})}
\newcommand{\bfi}{\begin{figure}[24]}
\newcommand{\efi}[1]{\caption{\label{#1}}\end{figure}}
\newcommand\fig[1]{Fig.~\ref{#1}}
\newcommand{\res}{respectively}
\newcommand\gl{\left}
\newcommand\gr{\right}

\newcommand{\CE}{{\cal E}}

\newcommand{\CM}{{\cal M}}

\newcommand{\CQ}{{\cal Q}}

\newcommand{\Ga}{\alpha}
\newcommand{\Gb}{\beta}

\newcommand{\Ge}{\varepsilon}

\newcommand{\Gf}{\phi}

\newcommand{\Gk}{\varkappa}
\newcommand{\Gl}{\lambda}
\newcommand{\Gn}{\eta}

\newcommand{\Gvk}{\varkappa}

\newcommand{\Go}{\omega}

\newcommand{\GF}{\Phi}

\newcommand{\az}[1]{Sect.$\!$ \ref{#1}}
\newcommand\D{\,\mathrm{d}}
\newcommand\I{\mathrm{i}}
\newcommand\E{\mathrm{e}}
\newcommand{\bexe}{\begin{quote}\begin{exercise}\inh}
\newcommand{\eexe}[1]{\label{#1}\end{exercise}\end{quote}}

\usepackage{graphics,graphicx}
\usepackage{color}
\begin{document}
{\large
\title{
Transition wave in a supported heavy beam
}}

\author{Michele Brun$^{a,b,*}$, Alexander B.  Movchan$^{a}$, Leonid I. Slepyan$^{c}$}
\date{
$^a$ {\em Department of Mathematical Sciences,}\\
{\em University of Liverpool, Liverpool, L69 7ZL, UK} \\
$^b${\em Dipartimento di Ingegneria Meccanica, Chimica e dei Materiali,} \\
{\em Universit\'a di Cagliari, Piazza d'Armi, I-09123 Cagliari, Italy} \\
$^c${\em School of Mechanical Engineering, Tel Aviv University\\
P.O. Box 39040, Ramat Aviv 69978 Tel Aviv, Israel}}


\maketitle

\vspace{10mm}\noindent
{\bf Abstract}
We consider a heavy, uniform, elastic beam rested on periodically distributed supports as a simplified model of a bridge. The supports are subjected to a partial destruction propagating as a failure wave along the beam. Three related models are examined and compared: (a) a uniform elastic beam on a distributed elastic foundation, (b) an elastic beam which mass is concentrated at a discrete set of points corresponding to the discrete set of the elastic supports and (c) a uniform elastic beam on a set of discrete elastic supports. Stiffness of the support is assumed to drop when the stress reaches a critical value. In the formulation, it is also assumed that,  at the moment of the support damage, the value of the `added mass', which reflects the dynamic response of the support,  is dropped too. Strong similarities in the behavior of the continuous and discrete-continuous models are detected.
Three speed regimes, subsonic, intersonic and supersonic, where the failure wave is or is not accompanied by elastic waves excited by the moving jump in the support stiffness, are considered and related characteristic speeds are determined. With respect to these continuous and discrete-continuous models, the conditions are found for the failure wave to exists, to propagate uniformly or to accelerate. It is also found that such beam-related transition wave can propagate steadily only at the intersonic speeds. It is remarkable that the steady-state speed appears to decrease as the jump of the stiffness increases.


\vspace{15mm}
$^*$ {\footnotesize Corresponding author.
Email: mbrun@unica.it;
tel.: +39 070 6755411, +44 [0] 151 7944002; \\
fax: +39 070 6755418, +44 [0] 151 794-406;
web-page: http://people.unica.it/brunmi/}

\vspace*{5mm}
\noindent Keywords: flexural waves, lattice system, Wiener-Hopf functional equations, fracture.


\section{Introduction}
In a system like a uniform or a periodic waveguide, it may happen that a localized damage causes a failure wave. Mechanically such process is similar to the phase transition, and it can occur if the wave is accompanied by a permanent energy release sufficient to overcome the associated energy barrier.  A well-known example is a falling row of dominoes. Many studies have been devoted to the related dynamic problems. Plane crushing waves were considered in Galin and Cherepanov (1966),  Grigoryan (1967), Slepyan (1968, 1977), Slepyan and Troyankina (1969), where, in particular, the uncertainty of a condition at the wavefront was discussed. It was then shown that uniqueness can be achieved in the framework of a structured material model where the speed-dependent wave resistance to the transition can be determined (see Slepyan, 2002). To describe the related phenomena a higher-order-derivative formulation for an elastic continuum was used  (see Truskinovsky (1994, 1997),
Ngan and Truskinovsky (1999), and the references therein), and a discrete chain model (Slepyan and Troyankina (1984, 1988)). Waves in discrete bistable chains were then studied in Puglisi and Truskinovsky (2000), Slepyan (2000, 2001), Balk et al. (2001a,b), Cherkaev et al. (2005), Slepyan et al. (2005), Vainchtein and Kevrekidis (2012). Localized transition wave in a two-dimensional lattice model was considered by Slepyan and Ayzenberg-Stepanenko (2004), while non linear effects have been studied with large scale atomistic models by Buelher et al. (2003).

With regard to a large-scale long-length construction, the failure wave may be supported by the gravity forces. In this connection we refer to papers by Ba$\check{z}$ant and Zhou (2002),  Ba$\check{z}$ant and Verdure (2007), Ba$\check{z}$ant et al. (2008), which contain a comprehensive analysis of the collapse wave progress in the nine-eleven disaster. A bridge on pillars or a suspended bridge, an overpass, long conveyers are examples of the constructions where the failure wave may propagate taking energy from the gravity forces.

In this paper, we examine some simplified models of such a construction considering the latter as a beam on a discrete supports and on a continuous elastic foundation, where the failure wave is that of a partial damage of the supports. Mechanically, these models differ from the above-mentioned ones by the existence of the subsonic and intersonic regimes of the failure wave propagation: no elastic wave is excited in the steady-state regime as far as it propagates in the subsonic speed range and there is such wave only behind the failure wavefront in the intersonic regime. Note that the elastic waves, if exist, create wave resistance to the failure wave. Possibilities of the wave radiation, in dependence on the failure wave speed, can be seen in the corresponding dispersion relations found for related structure in Brun et al. (2012).  Dispersion diagrams to compare are plotted in \fig{Fig10}.

\begin{figure}[!ht]

\centering
\vspace*{10mm} \rotatebox{0}{\resizebox{!}{10.cm}{%
\includegraphics[scale=0.5]{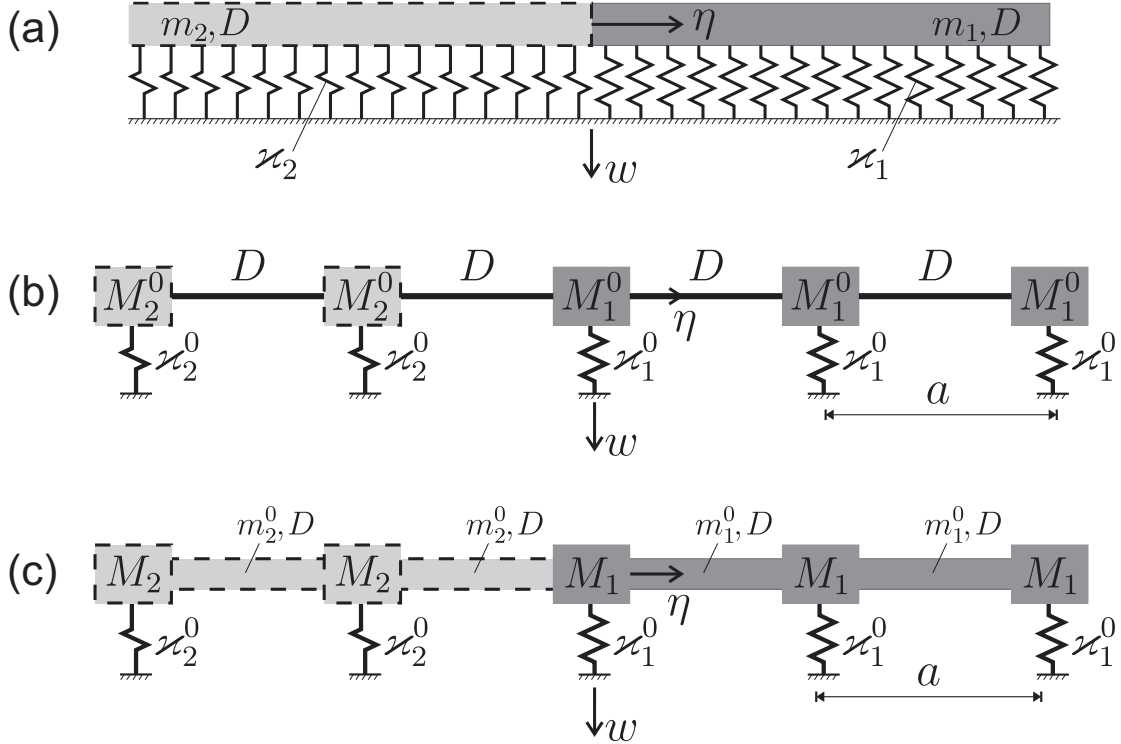}}}
 \caption{Supported beam models. Undamaged and damaged parts are indicated with subscripts $1$ and $2$, \res. (a) Beam on elastic foundation. The beam has mass 
 density $m_{1,2}=m^0_{1,2}+M_{1,2}/a$ and bending stiffness $D$; the foundation has stiffness per unit length $\Gvk_{1,2}=\Gvk_{1,2}^0/a$. (b) Discrete set of masses $M^0_{1,2}=m^0_{1,2}a+M_{1,2}$ connected horizontally by massless beams of bending stiffness $D$ and rested on elastic supports of stiffness $\Gvk^0_{1,2}$ placed at distance $a$. (c) Dynamic beam of density $m^0_{1,2}$, added mass $M_{1,2}$ and  bending stiffness $D$, on the set of discrete elastic supports of stiffness $\Gvk^0_{1,2}$ placed at distance $a$. The transverse displacement is denoted as $w$.}
    \label{Fig01}
\end{figure}

The supported beam model also differs by the fact that the failure wave speed can vary in a wide range depending on the structure and damage parameters, and it can be very small compared with that in the above-mentioned bistable models. It is remarkable that the speed limit appears as low as the jump of the support stiffness is large.
Three related models are examined: (a) a dynamic beam on a continuous elastic foundation, (b) a discrete set of masses rested on elastic supports and connected by massless beams and (c) a dynamic beam on the set of discrete elastic supports, \fig{Fig01}. Stiffness of the support is assumed to drop when the stress reaches a critical value. The failure wave is also considered under the condition that, at the moment of the support damage, the value of the `added mass', which reflects the dynamic response of the support, is dropped.
We have found the conditions for the failure wave to exists, to propagate uniformly or to accelerate. All speed regimes are considered, such where the failure wave is or is not accompanied by elastic waves excited by the jump in the stiffness and added mass. Limiting speeds are determined. We have shown that the effect of the change of the stiffness can essentially result in the failure wave speed limitations.

\begin{figure}[!ht]

\centering
\vspace*{10mm} \rotatebox{0}{\resizebox{!}{4.1cm}{%
\includegraphics[scale=0.5]{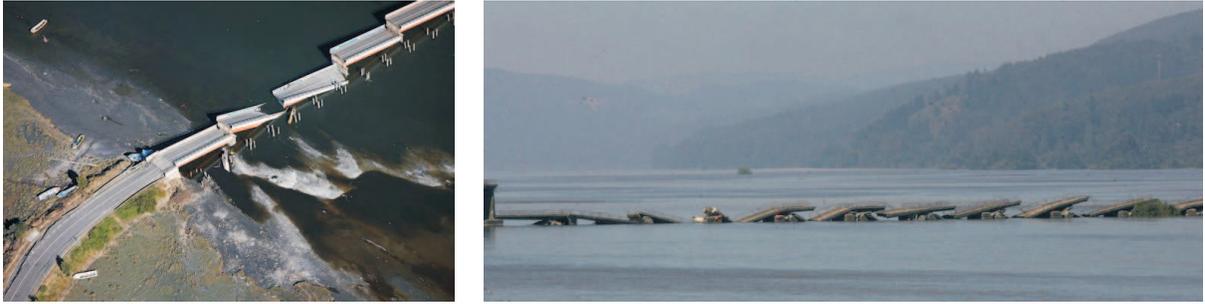}}}
 \caption{View of the destroyed ``Puente Viejo'' over the Biobio river, that links Concepcion and San Pedro de la Paz, 500 km south of Santiago, Chile, after the $8.8$ magnitude earthquake on February 27, 2010.  $\copyright$  Nicol\'as Piwonka, National Geographic.}
    \label{Fig02}
\end{figure}

In \fig{Fig02} we show an illustration of a bridge failed under a wave generated by an earthquake. A periodic pattern is clearly visible in the collapsed structures and it corresponds to damaged foundations of the bridges.


\section{Uniform continuous model:  a beam on an elastic foundation}
\label{Sect2}
We start with the simplest formulation of the structure as an infinite uniform elastic beam placed horizontally on a uniform elastic foundation as given in \fig{Fig01}a. The foundation is considered as massless; however, the corresponding `added mass' is assumed to be included in the beam per-unit-length mass, $m$. The foundation is initially stressed by the beam under gravity forces $mg$ ($g$ is the acceleration of gravity). We consider the failure wave as the propagation of partial damage of the foundation. Namely, we assume that the foundation stiffness drops at the wave front, $\Gn\equiv x-vt=0$, from its initial value, $\Gvk_1$ (ahead of the front) to $\Gvk_2<\Gvk_1$ (behind the front), while the per-unit-length load on the foundation reaches a critical value, $q_c$ at the front, $\Gn=0$. Also, this problem is considered assuming that a part of the added mass disappears simultaneously with the jump discontinuity of the foundation stiffness. In this case, $m=m_1$ ahead of the front and $m=m_2$ behind the front. Under the moving discontinuity, elastic waves can be excited propagating behind or/and ahead of the failure wave front; however, the static values of the beam vertical displacement are equal to $m_1g/\Gvk_1$ (far ahead of the front) and to $m_2g/\Gvk_2$ (far behind the front). This allows us to make some conclusions based on energy considerations.

\subsection{Energy considerations}\label{ecs}
\begin{figure}[!ht]

\centering
\vspace*{10mm} \rotatebox{0}{\resizebox{!}{10.cm}{%
\includegraphics[scale=0.5]{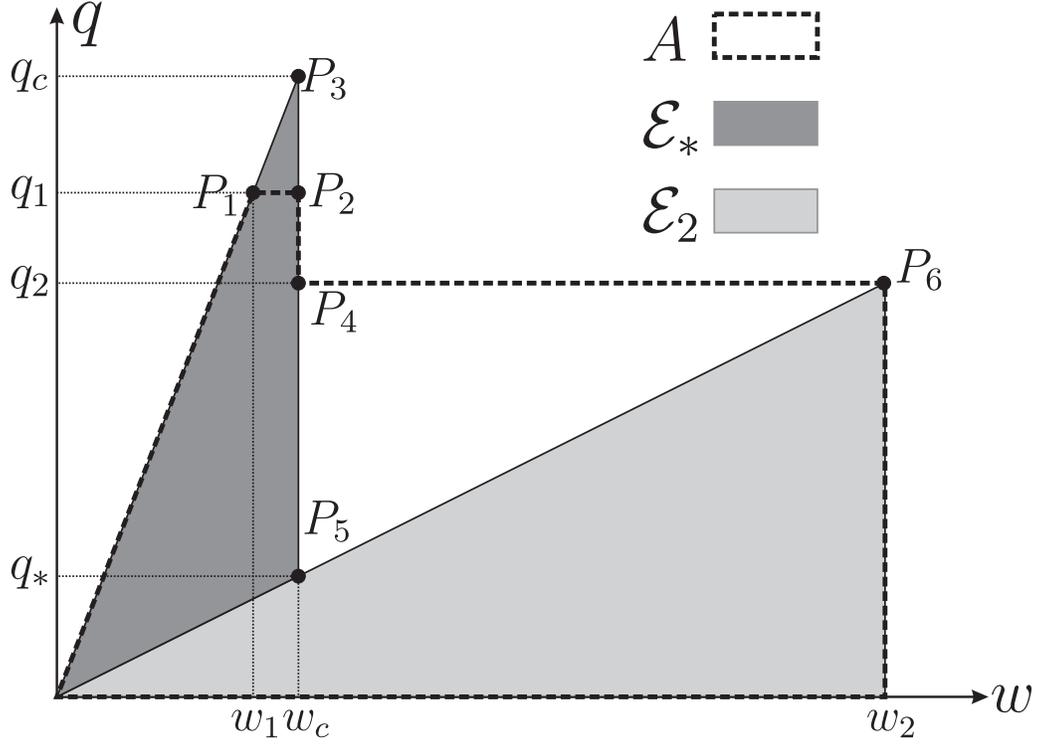}}}
 \caption{Load $q$ versus displacement $w$ diagram for the beam on elastic foundation; $q_1$ and $q_2$ are the gravitational loads for the undamaged and damaged structure of stiffness $\Gvk_1$ and $\Gvk_2$, \res, and $w_1=q_1/\Gvk_1$, $w_2=q_2/\Gvk_2$ are  the corresponding static displacements. The critical load and displacement at damage are $q_c$ and $w_c$, respectively, whereas $q_*$ is the load corresponding to the displacement $w_c$ of the damaged structure. $A$ is the work of the gravitational loads, $\CE_2$ is the static strain energy in the damaged structure at $\eta\rightarrow -\infty$ and $\CE_*$ is the energy loss at failure.
 The energy excess $\CE_0$ is the difference between the area of the two triangles $P_4P_5P_6$ and $P_1P_2P_3$ in the $(w,q)$ space.
 }
    \label{Fig03}
\end{figure}

Consider the loading diagram shown in \fig{Fig03}, where $q_{1,2}$, and $q_c$ are the initial, the final and the critical load at damage acting on the support, \res. The other values pointed in the figure are
 \beq q_* = \f{\Gvk_2}{\Gvk_1}q_c\,,~~~w_1= \f{q_1}{\Gvk_1}\,,~~~w_2= \f{q_2}{\Gvk_2}\,,~~~w_c= \f{q_c}{\Gvk_1}\,,\eeq{ecfsss1}
where $w_{1,2}$ is the beam static displacement at $\Gn = \pm \infty$
and $w_c$ is the critical displacement at $\eta=+0$.
It can be seen that the work of the gravity forces, the static strain energy density at $\Gn= - \infty$ and the energy loss due to the failure are
 \beq A=\f{1}{2}q_1w_1 +q_1(w_c-w_1)+q_2(w_2-w_c)\,,~~~\CE_2 =\f{1}{2}q_2w_2\,,~~~
 \CE_* = \f{1}{2}(q_c-q_*)w_c\,.\eeq{ecfsss2}
Since the energy of the beam itself in the steady-state regime is invariable, the energy excess per unit length is
 \beq \CE_0= A-\CE_2-\CE_*= \f{1}{2}\gl(\f{q_2^2}{\Gvk_2}-\f{q_1^2}{\Gvk_1}\gr)+\f{(q_1-q_2)q_c}{\Gvk_1}-\f{q_c^2}{2\Gvk_1}\gl(1-\f{\Gvk_2}{\Gvk_1}\gr)\,,\eeq{ecfsss3}
The energies are also shown in the load-displacement plane in \fig{Fig03}.
From this static value the energy radiated by elastic waves, if
exist, must be deducted. In this general case, for the uniform propagating failure wave the energy balance is
\beq\CE_0 - U_1(c_{1}/v - 1) - U_2(1-c_{2}/v) =0\,,\eeq{ecfsss4}
where $U_{1,2}$ and $c_{1,2}$ are the energy density and the group velocities of the waves at $\pm \Gn >0$.
Note that $U_1=0$ in the sub- and intersonic regimes and $U_2=0$ in the subsonic regime.

In this problem, there is a range of the failure wave speed, $0\le v < v_2$,  where no elastic wave can be radiated in steady-state motion.
The critical velocity $v_2$, indicated in \fig{Fig04}, is the resonant wave velocity corresponding to the lower dispersion curve. At this regime the phase and group velocities are equal. The value of $v_2$ is given the next section.
For this subsonic regime, we can conclude that the failure wave cannot propagate if $\CE_0<0$. In the opposite case, where there is a positive excess of the energy, the steady-state regime is impossible, and the failure wavefront move unsteadily spending the energy excess on the radiation of elastic waves, which arise under nonuniform motion. In the neutral case, $\CE_0=0$, any value of the speed in the subsonic range satisfies the energy balance. However, in the latter case, the system parameters are connected by a relation following from the equation $\CE_0=0$. In particular, if no loss of the mass is assumed, this relation is
 \beq w_c = w_c^*=\,w_1\sqrt{\Gvk_1/\Gvk_2}\,.\eeq{ecfsss5}
In the subsonic regime, the wave propagates nonuniformly if the real critical displacement is less than this value, otherwise it cannot propagate.
In the case ${\cal E}_0>0$ steady-state motion is possible at $v>v_2$, solution of \eq{ecfsss4} as shown in the example at the end of \az{Sect2}.

Note that the last term in Eq. \eq{ecfsss3}, the energy lost under the support damage, plays the role of the surface energy (or the effective surface energy) in fracture. The peculiarity of the considered problem is that the energy release under the support damage remains positive not only in the cases where there is no wave radiation, but also in a part of the intersonic speed range, and this occurs in the continuous model as well. In this latter case, the radiation decreases but not eliminates the total energy released under the `phase transition'.

\subsection{Flexural waves, normalization and dispersion relations}
\label{Sect2.2}
We based on the Bernoulli-Euler model for the beam and the Winkler model for the continuous foundation as in \fig{Fig01}a. Thus, the dynamic equation is
 \beq
 D\f{\p^4 w(x,t)}{\p x^4} + m \f{\p^2 w(x,t)}{\p t^2} + \Gvk w(x,t)= m g\,.
 \eeq{bee1}
where $w$ is the transversal displacement, \res,  $\Gvk$ is the foundation stiffness and $g$ is the acceleration of gravity. The foundation stiffness and possibly the density take different values ahead of and behind the wavefront. This continuous structure is assumed to be a limiting representation of the corresponding discrete-continuous system consisting of a continuous beam on a discrete periodic set of supports. Accordingly, the distributed mass density and support stiffness are assumed to be those averaged over the period, $a$
 \beq m=m^0 + M/a\,,~~~\Gvk = \Gvk^0/a\,,\eeq{d-cc1}
where $m^0$ is the beam mass density, $M$ is the added mass due to the supports and $\Gvk^0$ is the discrete support stiffness (see also \az{dcm}).

We consider the steady-state regime, assuming that the displacement depends only on $\Gn = x-vt$, where $v$ is the failure wave speed. The equation of motion \eq{bee1} becomes
 \beq
 D w(\Gn)^{IV} + m v^2 w''(\Gn) + \Gvk \,w(\Gn)= m g\eeq{bee2}
with
 \beq \Gvk=\Gvk_1~~(\Gn>0)\,,~~~\Gvk=\Gvk_2<\Gvk_1~~(\Gn<0)\,,\n m=m_1~~(\Gn>0)\,,~~m=m_2~~(\Gn<0)\,.
 \eeq{bee3}
In this equation, $v$ is considered as the input parameter. Ones the steady-state solution is obtained, it must be used to satisfy the failure condition
 \beq w(0)=w_c\,.\eeq{wfzetwc}
If the latter equation can be satisfied it serves for the determination of the speed $v$; otherwise, the steady-state formulation failed as it was discussed in \az{ecs}.

\begin{figure}[!ht]

\centering
\vspace*{10mm} \rotatebox{0}{\resizebox{!}{8.5cm}{%
\includegraphics
[scale=0.5]{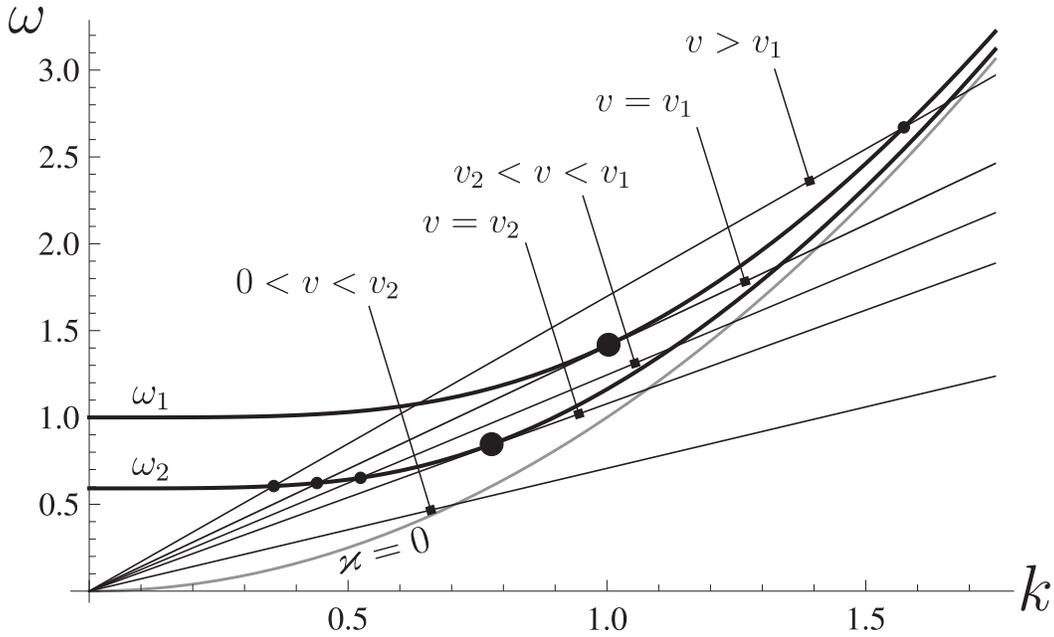}}}
 \caption{
 Dispersion diagrams for the waves in uniform beams on elastic foundation of stiffness $\Gvk$. The dispersion curves are shown for the undamaged foundation ($\omega_1(k)$ for stiffness $\Gvk=\Gvk_1$), the damaged one ($\omega_2(k)$ for $\Gvk=\Gvk_2$, with $\hat{\Gk}=\Gk_2/\Gk_1=0.36$) and the limiting case with $\Gvk=0$ (grey curve); no loss of mass is considered. Straight lines correspond to the different regimes of the failure wave speeds, $v=\Go/k$= const.
The group velocities $c_{1,2}=\D\Go_{1,2}/\D k$ coincide with the speed $v$ at the resonant regimes, where $v=v_{2}=\sqrt{2} (\hat{\Gvk})^{1/4}$ and $v=v_1=\sqrt{2}$.
The $(\Go, k)$-points corresponding to the radiated waves are marked by small bullets; such resonant points are marked by larger bullets. The  non-dimensional variables are used corresponding to the length and time units introduced in \eq{bee5a}.
}
\label{Fig04}
\end{figure}

We introduce the natural length and time units as
\beq
\xi=\left(D/\Gvk_1\right)^{1/4}\!\!,~~~\tau=\sqrt{m_1/\Gvk_1}\,.\eeq{bee5a}
After the normalization
 \beq
 (x,\Gn,w)=\xi(\tilde{x},\tilde{\Gn},\tilde{w})\,,~~~(t,\omega^{-1})=\tau(\tilde{t},\tilde \omega^{-1})\,, ~~~(v,c_{1,2})=\xi/\tau(\tilde{v},\tilde{c}_{1,2})\,,~~~\n
 k=\tilde{k}/\xi\,,~~~g=(\xi/\tau^2)\tilde{g} \,,~~~
 (\CE_0,\CE_2,\CE_*,A,U_1,U_2)=\Gvk_1\xi^2(\tilde\CE_0,\tilde\CE_2,\tilde\CE_*,\tilde A,\tilde U_1,\tilde U_2)\,,
 \eeq{bee5}
Eqs. \eq{bee2} become
 \beq
\tilde{w}^{IV}(\Gn) +\tilde v^2 \,\tilde w''(\Gn) + \tilde{w}(\eta)= \tilde{g}~~~(\Gn>0)\,,\n
\tilde{w}^{IV}(\Gn) +\hat{m}\, \tilde v^2 \,\tilde w''(\Gn) + \hat{\Gvk}\,\tilde{w}(\eta)= \hat m \tilde{g}~~~(\Gn<0)\,,
\eeq{bee6}
where $\hat{m}=m_2/m_1$ and $\hat{\Gvk}=\Gvk_2/\Gvk_1$.
Separating the initial static displacement of the undamaged structure we implement the substitution $\tilde w=\bar w +\tilde g$. Then, the above equation takes the form
\beq
\bar w^{IV}(\Gn) +\tilde v^2 \,\bar w''(\Gn) + \bar w(\eta)= 0~~~(\Gn>0)\,,\n
\bar w^{IV}(\Gn) + \hat m \, \tilde v^2 \,\bar w''(\Gn) + \hat\Gvk \bar w(\eta)= \CQ~~~(\Gn<0)\,,
\eeq{bee4}
where $\CQ = \tilde g \, (\hat m - \hat \Gvk)$.

In the following, the notations ``tilde'' and ``bar'' are omitted for ease of notation
and we consider the case $m_1=m_2$. The displacements referred to in the following text are normalised dynamic perturbations with respect to the corresponding static values.

The corresponding dispersion relations are
\begin{eqnarray}
\omega_1 = \pm\sqrt{1+ k^4}\,,  ~~~ (\eta>0)\,,~~~\omega_2 =\pm\sqrt{\hat{\Gvk}+ k^4}~~~(\Gn<0)\,,
\eeq{bee7}
where $k$ is the wave number.

We consider three ranges of the velocity, $v$, corresponding to subsonic, intersonic and  supersonic speeds. The boundaries separating these ranges correspond to equalities $v=\Go/k = c_{1,2}=\D\Go_{1,2}/\D k$ ($c_{1,2}$ are the group velocities). We find that
 the three ranges are identified as follows:
\begin{itemize}
\item $~~~$ \emph{subsonic} range: $~~~~~$ $0 \leq v <  v_2=\sqrt{2} (\hat{\Gvk})^{1/4} $\,,
\item $~~~$ \emph{intersonic} range: $~~~$$v_2< v <  v_1=\sqrt{2}$\,,
\item $~~~$ \emph{supersonic} range: $~~~$ $v>v_1$.
\end{itemize}
Dispersion diagrams \eq{bee7} and sonic regimes are shown in \fig{Fig04}. Elastic wave radiation at the failure wave front $\eta=0$ corresponds to the intersection points of the dispersion relations \eq{bee7} with the propagating wave rays, $\omega=v\,k$, and the radiated wave propagates at $\Gn<0$ if $v>c_{2}$ and at $\Gn>0$ if $v<c_{1}$.

\subsubsection{Subsonic regime}
Although, as follows from the energy considerations, the only `neutral case' correspond to the steady-state regime in this speed range, we consider
the subsonic regime in detail.
We show how the deformed shape of the beam in a vicinity of the transition point depends on the speed. Eqs. \eq{bee6} yield
\beq w(\eta)= \E^{-\Ga_1 \Gn}(A_1\cos \Gb_1\Gn +B_1\sin \Gb_1\Gn)~~~(\Gn>0)\,,\n
w(\eta)=  \E^{\Ga_2 \Gn}(A_2\cos \Gb_2\Gn +B_2\sin \Gb_2\Gn) +\CQ/\hat\Gvk~~~(\Gn<0)\,,\n
 \Ga_1=\f{1}{2}\sqrt{2-v^2}\,,~~~ \Gb_1=\f{1}{2}\sqrt{2+v^2}\,,\n
\Ga_2=\f{1}{2}\sqrt{2\sqrt{\hat{\Gvk}}-v^2}\,,~~~ \Gb_2=\f{1}{2}\sqrt{2\sqrt{\hat{\Gvk}}+v^2}
\,,\eeq{bee8}
where the constants, $A_1, ..., B_2$, defined by the continuity conditions with respect to the displacement and its first three derivatives at $\Gn=0$, are
 \beq
A_1 = \f{\CQ}{\sqrt{\hat\Gvk} (1+\sqrt{\hat\Gvk})}\,,~~~  A_2 = - \f{ \CQ }{\hat\Gvk (1+\sqrt{\hat\Gvk})}\,,\n
B_1= - \f{\sqrt{2 - v^2} ( 1 + \sqrt{\hat\Gvk} ) - 2  \sqrt{2\sqrt{\hat\Gvk} - v^2}   }{\sqrt{\hat\Gvk}\sqrt{2 + v^2} ( 1- \hat\Gvk ) }\CQ \,,\n
B_2 = \f{\sqrt{2 \sqrt{\hat \Gvk} - v^2} ( 1 + \sqrt{\hat \Gvk} ) - 2 \sqrt{\hat \Gvk}  \sqrt{2 - v^2}   }{\hat\Gvk\sqrt{2\sqrt{\hat \Gvk} + v^2} ( 1 - \hat \Gvk ) }\CQ \,.
 \eeq{bee9}

The displacement at the transition point
 \beq w(0) =  \f{\CQ}{\sqrt{\hat\Gvk} (1+\sqrt{\hat\Gvk})}=\gl( \sqrt{\f{ \Gvk_1}{\Gvk_2}}-1\gr)g\eeq{bee10}
is independent of the speed as it should be in the subsonic speed range. It corresponds to the energy balance relation \eq{ecfsss5} (note that in \eq{bee10} we have separated the  initial non-dimensional  displacement equal to $g$). This regime can exist if the critical value of the displacement coincide with $w(0)$ given by this relation. The plots of $w(\Gn)$ and of $w(0)$ as a function of $\hat{\Gvk}$ are presented in \fig{Fig05}.
In connection with energy considerations given in \az{ecs}, we can see from \fig{Fig05}b that, independently on the velocity $v\le \sqrt{2} (\hat \Gvk)^{1/4}$, the wave propagate nonuniformly if $\Gvk_2/\Gvk_1<(w_c/g+1)^{-2}$ (in normalized coordinates) and it cannot propagate otherwise. In this regime the critical displacement for dynamic propagation coincides with the static displacement.

\begin{figure}[!ht]

\centering
\vspace*{10mm} \rotatebox{0}
{\resizebox{!}{5.cm}{\includegraphics[scale=0.5]{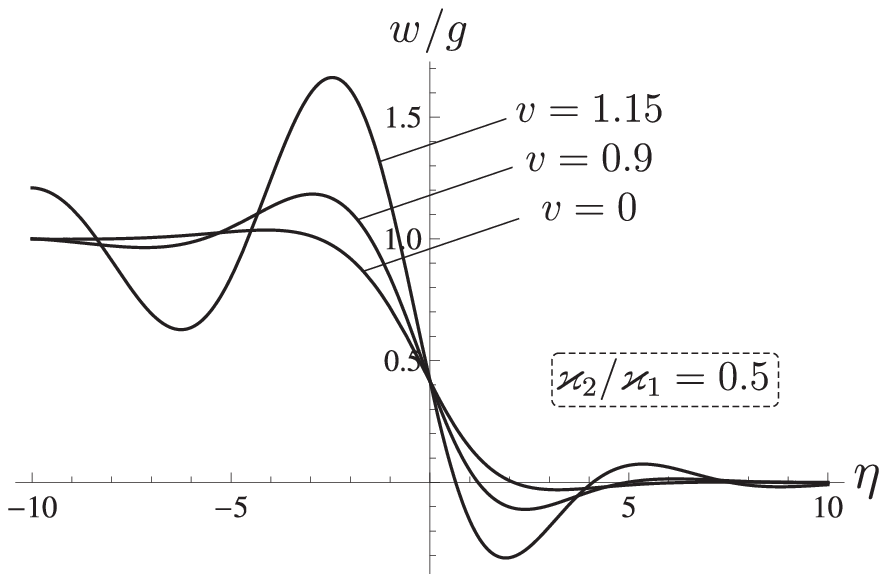}}} ~~~~
{\resizebox{!}{5.cm}{\includegraphics[scale=0.5]{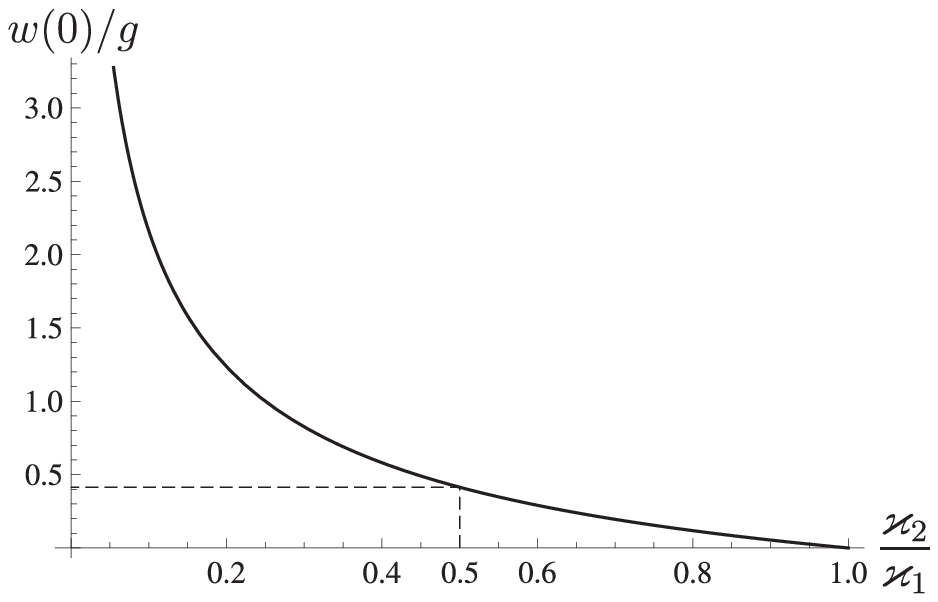}}}\\
(a) ~~~~~~~~~~~~~~~~~~~~~~~~~~~~~~~~~~~~~~~~~~~~~~~~~~~~~~~~~(b)
\caption{The beam on continuous elastic foundation. Subsonic speed regime ($v<\sqrt{2}{\hat \Gvk}^{1/4}$): (a) Displacement profile $w(\Gn)$, in the static case ($v=0$) and at velocities $v=0.9,1.15$, for $\hat \Gvk=0.5$ ($v_2=1.189$). (b) Displacement $w/g$ at $\eta=0$ as a function of the stiffness ratio $\hat \Gvk=\Gvk_2/\Gvk_1$. Dashed lines indicate the case shown in \fig{Fig05}a.}
    \label{Fig05}
\end{figure}

\subsubsection{Intersonic regime}
This regime is characterized by a sinusoidal wave propagating behind the transition point. Among two waves satisfying the equation we choose the wave with group velocity  below  the failure wave velocity (see \fig{Fig04}), that is we keep the wave excited by the transition (the other propagating from minus infinity could correspond to a remote source). Thus
 \beq w(\Gn) =  \E^{-\Ga_1 \Gn}(A_1\cos \Gb_1\Gn +B_1\sin \Gb_1\Gn)~~~(\Gn>0)\,,\n
w(\Gn)= A_2\cos \Gb_2\Gn +B_2\sin \Gb_2\Gn +\CQ/\hat{\Gvk}~~~(\Gn<0)\,,\n
 \Ga_1=\f{1}{2}\sqrt{2-v^2}\,,~~~ \Gb_1=\f{1}{2}\sqrt{2+v^2}\,,\n
\Gb_2=\sqrt{v^2/2 - \sqrt{v^4/4 -\hat{\Gvk}}}
\,,\eeq{bee11}
where the constants, $A_1, ..., B_2$, are defined by the same continuity conditions as above for the subsonic case; they are
 \beq
A_1 =\f{ v^2 - 2\hat\Gvk -\sqrt{v^4 - 4 \hat \Gvk} }{2 (1-\hat \Gvk) \hat\Gvk} \CQ\,,~~~  B_1= -\sqrt{2 - v^2} \,\f{v^2 + 2 \hat \Gvk - \sqrt{v^4 - 4 \hat\Gvk} }{2 \sqrt{2 + v^2} (1-\hat\Gvk) \hat \Gvk} \CQ\,,\n
A_2 = -\f{2-v^2 + \sqrt{v^4 - 4 \hat\Gvk}}{2 (1-\hat \Gvk) \hat \Gvk}\CQ\,,~~~
B_2 =  -\sqrt{2 - v^2} \f{\sqrt{v^2 - \sqrt{v^4 - 4 \hat \Gvk}}}{\sqrt{2}(1-\hat \Gvk) \hat \Gvk} \CQ\,.
\eeq{bee12}

In particular,
 \beq w(0) = \f{v^2 - 2 \hat\Gvk - \sqrt{v^4 - 4 \hat \Gvk}}{2 \hat \Gvk}g\,.\eeq{bee13}
Since $w(0)$ depends on $v$ and this allows us to satisfy the condition \eq{wfzetwc}. The displacement profiles $w(\eta)$ for different values of the speed $v$ are presented in Fig. \ref{Fig06}a.

\begin{figure}[!ht]

\centering
\vspace*{10mm} \rotatebox{0}
{\resizebox{!}{5.cm}{\includegraphics[scale=0.5]{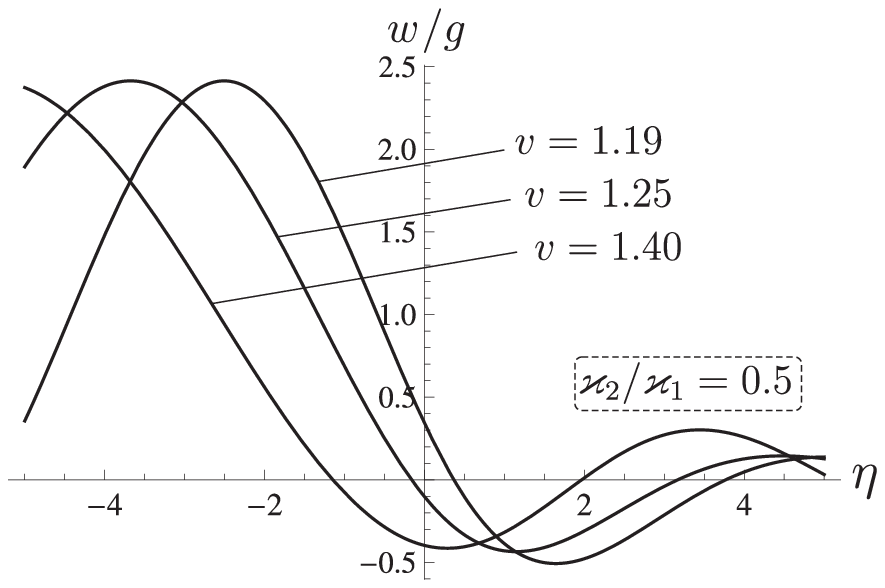}}}~~~~
{\resizebox{!}{5.cm}{\includegraphics[scale=0.5]{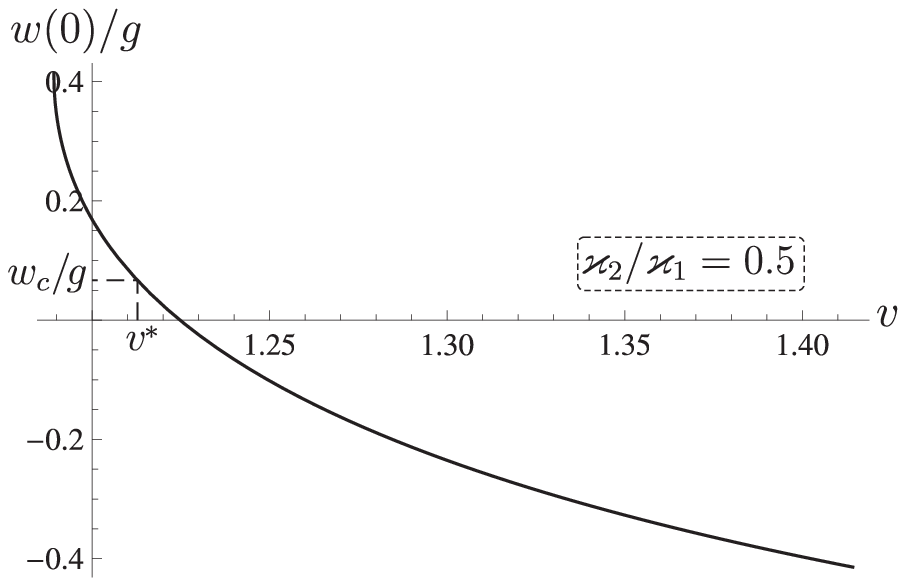}}}\\
(a) ~~~~~~~~~~~~~~~~~~~~~~~~~~~~~~~~~~~~~~~~~~~~~~~~~~~~~~~~~(b)
 \caption{The beam on an elastic foundation. Intersonic regime ($\sqrt{2}{\hat \Gvk}^{1/4}<v<\sqrt{2}$):
 (a) Displacement profile $w(\Gn)$ at different velocities, $v=1.19,1.25,1.40$, for $\hat \Gvk=0.5$.
 (b) Displacement at $\eta=0$ as a function of velocity for the ratio $\hat \Gvk=0.5$ ($1.189<v<1.414$); the critical displacement $w_c$ is attained at $v=v^*$.}
    \label{Fig06}
\end{figure}

In Fig. \ref{Fig06}b we also plot the value of $w(0)$ as a function of velocity $v$ for $\hat \Gvk=0.5$.
The steady-state propagation speed $v$ corresponds to equality $w(0) = w_c>0$ when $\Gvk_2/\Gvk_1<(w_c/g+1)^{-2}$. We note that
such a speed $v$ depends on the safety factor $w_c/g+1$ (expressed in term of non-dimensional values).

As can be seen in \fig{Fig06}b the displacement, $w(0)$, monotonically decreases as the speed, $v$, increases from the critical value, $v=v_2=\sqrt{2} (\hat{\Gvk})^{1/4}$, and becomes negative when the speed exceeds a value. These negative values correspond to a hypothetic case where the energy lost at the moment of the damage is less than its real minimal value, that is the critical load is less than the initial static load, $q_c < mg$.

The critical displacement $w(0)$ versus $v$ is shown in Fig. \ref{Fig07}a for different values of $\hat \Gvk$, while the total speed bound $w(0)=0$ is given in the $(\hat\Gvk,v)$ space in Fig. \ref{Fig07}b.
The steady propagation of the fault in the intersonic regime is not forbidden when $w(0)>0$, which implies that the velocity $v$ satisfies the following constraint
\beq
\sqrt{2} \hat \Gvk^{1/4} < v  < \sqrt{1+\hat \Gvk}.
\eeq{v_constr}
It is also noted that for the intersonic velocities within the interval $(\sqrt{1+\hat \Gvk}, \sqrt{2})$ the failure wave propagation is not possible.

\begin{figure}[!ht]

\centering
\vspace*{10mm} \rotatebox{0}
{\resizebox{!}{5.cm}{\includegraphics[scale=0.5]{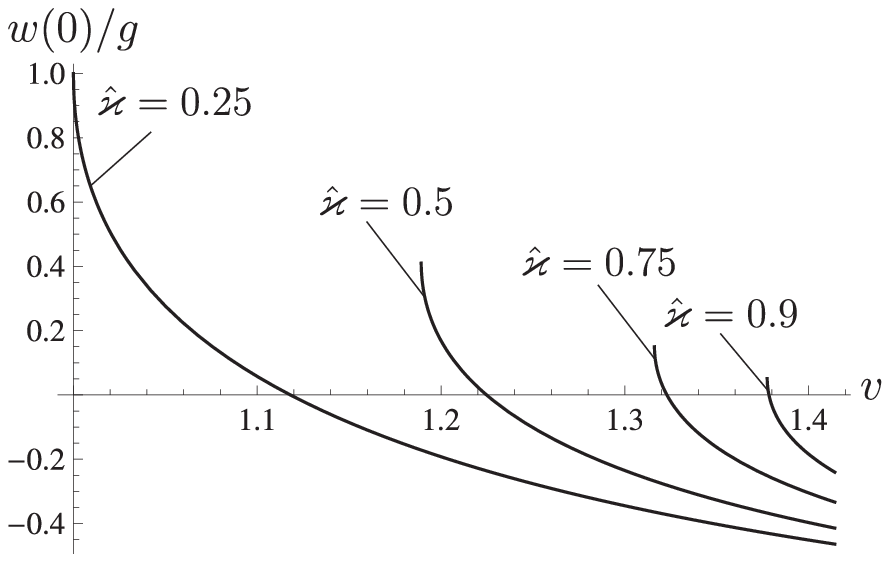}}}~~~~
{\resizebox{!}{5.cm}{\includegraphics[scale=0.5]{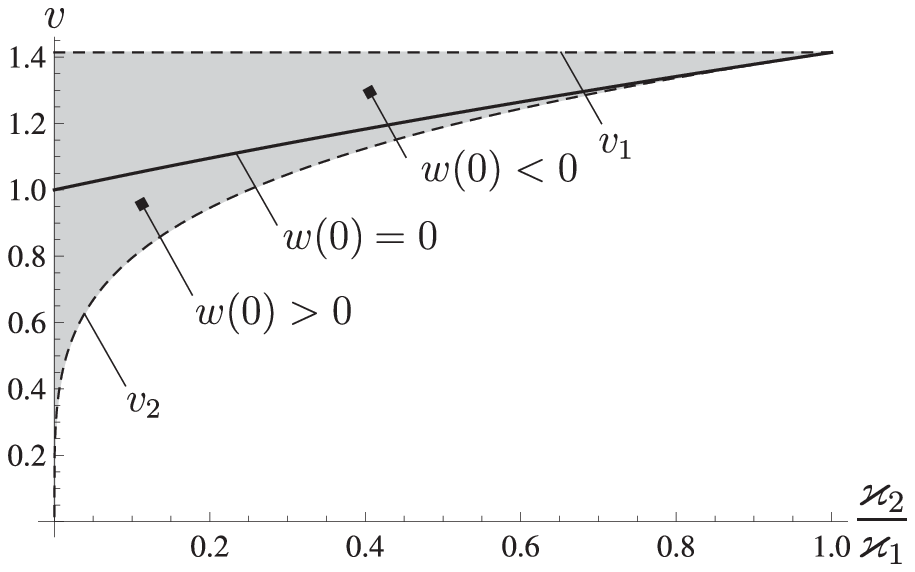}}}\\
(a) ~~~~~~~~~~~~~~~~~~~~~~~~~~~~~~~~~~~~~~~~~~~~~~~~~~~~~~~~~(b)
\caption{The beam on an elastic foundation. Intersonic regime ($\sqrt{2}{\hat \Gvk}^{1/4}<v<\sqrt{2}$). (a) Displacement at $\eta=0$ as a function of velocity for the ratios $\hat \Gvk=0.25,\,0.5,\,0.75,\,0.9$.
(b) Total bound $w(0)=0$ at different values of the velocity $v$ and the support stiffness contrast $\hat\Gvk$.
The grey area shows the intersonic regime, bounded by the dashed lines $v_2=\sqrt{2}(\hat\Gvk)^{1/4}$ and $v_1=\sqrt{2}$. Steady failure propagation is possible when $w(0)>0$.}
\label{Fig07}
\end{figure}

\subsubsection{Supersonic regime}
Finally, we address the case when the velocity of the transition front $ v > \sqrt{2}$. In this case, elastic waves are generated in both regions $\eta>0$ and $\eta<0$.  The displacement takes the form
 \beq w(\Gn) =  A_1 \cos \Gb_1\Gn +B_1 \sin \Gb_1\Gn~~~(\Gn>0)\,,\n
 w(\Gn)= A_2\cos \Gb_2\Gn +B_2 \sin \Gb_2\Gn +\CQ/\hat{\Gvk}~~~(\Gn<0)\,,\n
 \Gb_1=\sqrt{v^2/2 +\sqrt{v^2/4 -1}}\,,~~~\Gb_2=\sqrt{v^2/2 - \sqrt{v^2/4 -\hat{\Gvk}}}
 \,,\eeq{bee11ssr1}
where the group velocity of the elastic wave ahead of the transition front is greater than the speed of the latter (see \fig{Fig04}), and the constants defined by the continuity conditions are
 \beq
A_1  = - \frac{v^2-\sqrt{v^4-4\hat\Gvk}}{\sqrt{ v^4-4} + \sqrt{
   v^4 - 4 \hat \Gvk}}\,\,\frac{\CQ}{\hat \Gvk}\,,~~~B_1 =0 \,,\n
A_2 = - \frac{v^2+\sqrt{v^4-4}}{\sqrt{ v^4-4} + \sqrt{
   v^4 - 4 \hat \Gvk}}\,\,\frac{\CQ}{\hat \Gvk}\,,~~~B_2 =0\,.\eeq{bee11ssr1AB}
Thus, \begin{equation}
w(0)= - \frac{v^2-\sqrt{v^4-4\hat\Gvk}}{\sqrt{ v^4-4} + \sqrt{
   v^4 - 4 \hat \Gvk}}\,\,\frac{1-\hat \Gvk}{\hat \Gvk}g<0.
\label{bee14w}
\end{equation}
Since the transition can occur when $w(0)>0$ this evidences that the wave cannot propagate at supersonic velocities.
Thus, the absolute maximum of the failure wave speed is $\sqrt{1+\hat \Gvk}$.

\begin{figure}[!ht]

\centering
\vspace*{10mm} \rotatebox{0}
{\resizebox{!}{8.cm}{\includegraphics[scale=0.5]{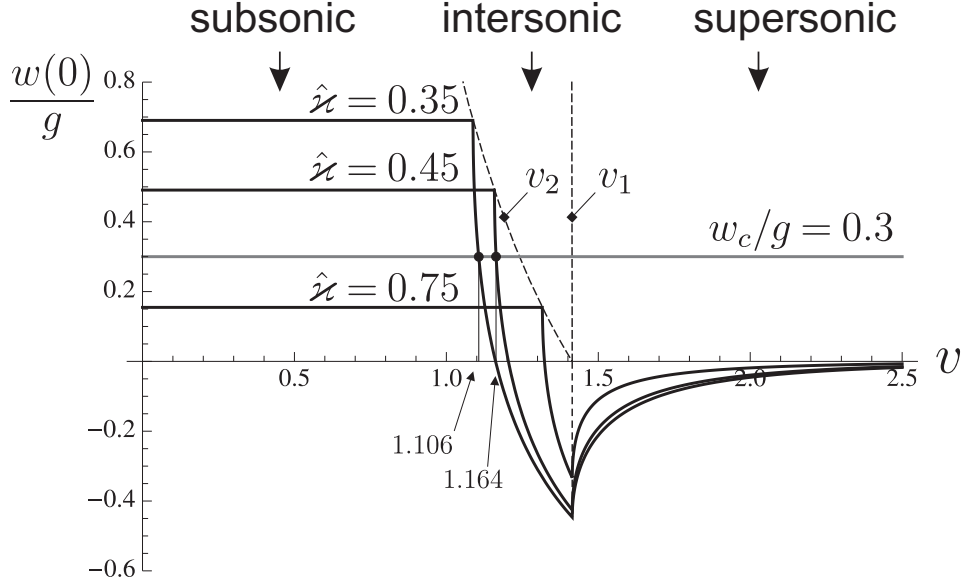}}}
 \caption{The beam on an elastic foundation. Displacement at the transition point is given in black continuous lines as a function of velocities in the subsonic, intersonic and supersonic regimes for $\Gvk_1/\Gvk_2=0.35,0.45,0.75$. Grey line indicates the critical displacement $w_c/g=0.3$ and black circles correspond to the steady-state propagation of the fault at $v^*=1.106,1.164$.}
    \label{Fig08}
\end{figure}

We conclude this section on the beam on elastic foundation with the results shown in \fig{Fig08}. The displacement $w(0)$ is plotted as a function of the failure velocity $v$ for the stiffness ratios $\hat\Gvk=0.35,\,0.45,\,0.75$. It is possible to distinguish the subsonic regime (where $w(0)$ is constant), the intersonic regime (where $w(0)$ decreases with the increase of $v$) and the supersonic regime (where $w(0)$ is always negative), delimited by the critical velocities $v_2=\sqrt{2}\hat\Gvk^{1/4}$ and $v_1=\sqrt{2}$. At the critical displacement $w_c/g=0.3$ corresponds a maximum ratio $\hat\Gvk=0.592$ for the fault propagation. Therefore, propagation is not possible for $\hat\Gvk=0.75$, which is also in agreement with the fact that in \eq{ecfsss3} the energy excess $\CE_0=-0.357(1-\hat\Gvk)  
g^2/2<0$. For $\hat\Gvk=0.35,\,0.45$, where $\CE_0$ is positive, steady-state propagation is given by the intersection points $w(0)=w_c$ at $v=1.106, 1.164$ (see Eq. \eq{bee13}), respectively; two values which satisfy the constraint \eq{v_constr}. It is trivial to verify that steady-state velocities satisfy energy balance \eq{ecfsss4} where
\begin{equation}
U_1=0,\quad U_2=\frac{1-\hat\Gvk}{4\hat\Gvk^2}\left(v^4- v^2\sqrt{v^4-4\hat\Gvk}\right),
\quad1-\frac{c_{2}}{v}=\frac{\sqrt{v^4-4\hat\Gvk}}{v^2}.
\label{eqn015}
\end{equation}

\section{Discrete-continuous model}\label{dcm}

A uniform elastic beam of the mass density $m^0$ is assumed to be placed on concentrated elastic supports at $x=n a, a>0, n=0, \pm 1, ...$ (see Fig. \ref{Fig01}c). We denote the stiffness, $\Gvk^0$, of the intact and damaged supports by $\Gk_1^0$ and $\Gk_2^0$, \res. Also the ``added mass'' $M$ proportional to $a$ is assumed to be placed at the junction points corresponding to positions of elastic supports.
Thus,
 \beq \Gk^0=\Gk_1^0\,,~~ M=M_1\,,~~m^0=m^0_1  ~~~~(\mbox{before the support damage,}\,\Gn > 0)\,,\n
 \Gk^0=\Gk_2^0<\Gvk_1^0\,,~~ M=M_2\,,~~m^0=m^0_2~~~(\mbox{after the support damage,}\,\Gn < 0)\,.\eeq{ssaamds1}
The supported cross-sections have the coordinates $x=n a$. The steady-state regime is considered
in the sense that the displacement $w$ at $x=n a$  is a function of $\Gn=n a-vt$,
whereas the displacement of the beam between the supports is a function of two variables: $w=w(x,\Gn)$. Thus, $w(\Gn)=w(n a,\Gn)$.

The dynamic equations for the structure before the damage 
are
 \beq D\f{\p^4 w(x,\Gn)}{\p x^4} + m^0_1 \f{\p^2 w(x,\Gn)}{\p t^2} = m^0_1 g ~~~(x\ne an)\,;\n
 M_1 \f{\p^2 w(\Gn)}{\p t^2} +\Gk_1^0w(\Gn) -Q^+(\Gn)+Q^-(\Gn) = M_1g\,,\n \CM^+(\Gn)-\CM^-(\Gn)=0~~~(x = an)
\,,\eeq{tbabexe1}
where $D$ is the bending stiffness, $m^0_1$ is the beam mass per unit length and $g$  is the gravity acceleration; $Q^\pm(\Gn) $ and $\CM^\pm(\Gn)$ are the transverse force and the bending moment at the right (superscript +) and at the left (superscript $-$) of the junction point $x= n a$, respectively  (see \fig{Fig09}, panel I).

\begin{figure}[!ht]

\centering
\vspace*{10mm} \rotatebox{0}
{\resizebox{!}{6.cm}{\includegraphics[scale=0.5]{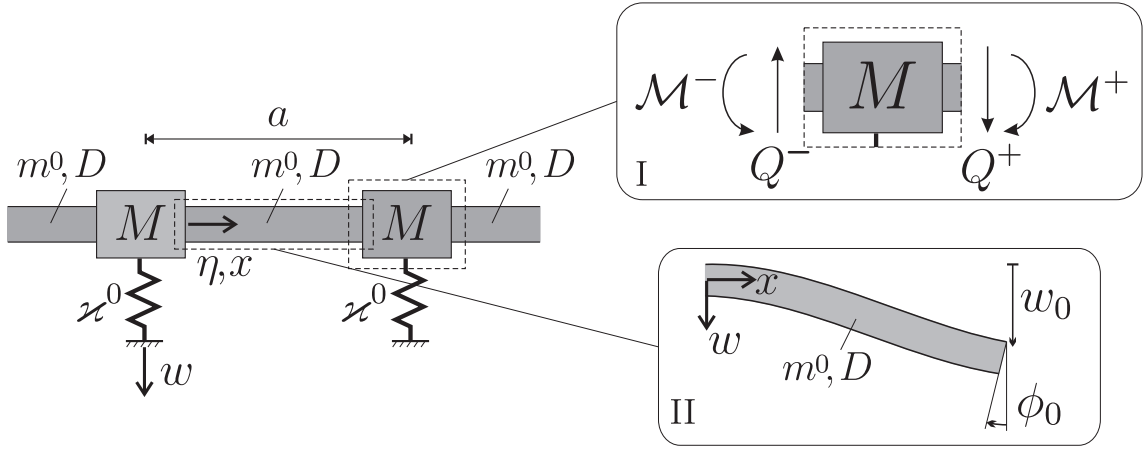}}}
 \caption{Inertial beam on elastic supports distributed at distance $a$. Panel I: configuration of the concentrated mass M. Panel II: auxiliary solution configuration for the beam.}
    \label{Fig09}
\end{figure}

We introduce the same length and time units  $\xi=\left(D/\Gvk_1\right)^{1/4}$, $\tau=\sqrt{m_1/\Gvk_1}$ as in \eq{bee5a}, with $\Gvk_1=\Gvk_1^0/a$, and the mass density unit $m_1=m^0_1+M_1/a$.
In these terms, the normalized quantities denoted by tilde are
 \beq
 (x, a, \Gn, w) = \xi (\tilde{x}, \tilde{a}, \tilde{\Gn}, \tilde{w}) , \, M_{1,2} = m_1 a\tilde{M}_{1,2}\,, \n
~~ (t,\omega^{-1})=\tau(\tilde{t},\tilde \omega^{-1})\,,~~~v=(\xi/\tau)\tilde{v}\,,~~~ g=(\xi/\tau^2)\tilde{g}\,,\n
 Q(x,\Gn) = -D\f{\p^3 w(x,\Gn)}{\p x^3} = \f{D}{\xi^2}\tilde{Q}(\tilde{x},\tilde{\Gn})\,,~~~\tilde{Q}(\tilde{x},\tilde{\Gn})=-\f{\p^3 \tilde{w}(\tilde{x},\tilde{\Gn})}{\p \tilde{x}^3}\,,\n
 \CM(x,\Gn)=D\f{\p^2 w(x,\Gn)}{\p x^2} =\f{D}{\xi}\tilde{\CM}(\tilde{x},\tilde{\Gn})\,,~~~\tilde{\CM}(\tilde{x},\tilde{\Gn})=\f{\p^2 \tilde{w}(\tilde{x},\tilde{\Gn})}{\p \tilde{x}^2}\,.\eeq{tbabexe2}
The equations of motion \eq{tbabexe1} become
 \beq
 \f{\p^4 \tilde{w}(\tilde{x},\tilde{\Gn})}{\p \tilde{x}^4} +  \frac{m^0_1}{m_1}\f{\p^2 \tilde{w}(\tilde{x},\tilde{\Gn})}{\p \tilde{t}^2} = \frac{m^0_1}{m_1}\tilde{g}~~~(x\ne an)\,; \label{tbabexe2a} \\
  \tilde{M}_1  \f{\p^2 \tilde{w}(\tilde{\Gn})}{\p \tilde{t}^2} +\tilde{w}(\Gn) -\f{1}{\tilde{a}}\gl[\tilde{Q}^+(\tilde{\Gn})-\tilde{Q}^-(\tilde{\Gn})\gr] =  \tilde{M}_1 \tilde{g}\,,~~ \tilde{\CM}^+(\tilde{\Gn})-\tilde{\CM}^-(\tilde{\Gn})=0~~~(x=an)
\,.\eeq{tbabexe2c}
Recall that in the damaged region, $\eta < 0$, the values  $\Gk_1^0, M_1$, $m^0_1$ are replaced by $\Gk_2^0<\Gk_1^0,$, $M_2$ and $m^0_2$.
For this case, equation \eq{tbabexe2a} and the first equation in \eq{tbabexe2c} take the form
\begin{eqnarray}
\f{\p^4 \tilde{w}(\tilde{x},\tilde{\Gn})}{\p \tilde{x}^4} +  \frac{m^0_2}{m_1}\f{\p^2 \tilde{w}(\tilde{x},\tilde{\Gn})}{\p \tilde{t}^2} = \frac{m^0_2}{m_1}\tilde{g}~~~(x\ne an)\,;\\
\tilde{M}_2  \f{\p^2 \tilde{w}(\tilde{\Gn})}{\p \tilde{t}^2} + \f{\Gvk_2}{\Gvk_1} \tilde{w}(\tilde\Gn) -\f{1}{\tilde{a}}\gl[\tilde{Q}^+(\tilde{\Gn})-\tilde{Q}^-(\tilde{\Gn})\gr] =  \tilde{M}_2 \tilde{g}\,.
\label{tbabexe2b}
\end{eqnarray}
In the following the symbol ``tilde'' is omitted for convenience.

We now separate the initial static state by the substitution
\beq w(x,\Gn) =g +\f{m^0_1}{24\,m_1}(x-an)^2(x-a(n+1))^2 g +\bar{w}(x,\Gn)\,,\eeq{tbabexe3}
where the first term on the right-hand side represents the static displacement at $x=na$ corresponding to the positions of  the supporting pillars,
whereas the second one stands for the relative displacement of the beam with respect to the supporting pillars and $\bar w$ is the normalised dynamic perturbation.
Correspondingly,
\beq
Q^\pm(\eta)=\pm\frac{m^0_1}{m_1}\frac{ g\,a}{2}+\bar Q^\pm(\eta), \qquad
\CM^\pm(\eta)=\frac{m^0_1}{m_1}\frac{g\,a^2}{12}+\bar \CM^\pm(\eta).
\eeq{tbabexe3b}
We obtain that for $\Gn > 0$
\beq  x \ne an: ~~~\f{\p^4  \bar{w}(x,\Gn)}{\p x^4} + \f{m^0_1}{m_1} \f{\p^2  \bar{w}(x,\Gn)}{\p t^2} =0\,;\n
  x=an:~~~M_1 \f{\p^2  \bar{w}(\Gn)}{\p t^2} +\bar{w}(\Gn) -\f{1}{a}\gl[\bar{Q}^+(\Gn)-\bar{Q}^-(\Gn)\gr] =0\,,\n
 \bar{\CM}^+(\Gn)-\bar{\CM}^-(\Gn)=0\,,
\eeq{tbabexe4}
and for $\Gn<0$
\beq x\ne an: ~~~ \f{\p^4  \bar{w}(x,\Gn)}{\p x^4} + \f{m^0_2}{m_1}  \f{\p^2  \bar{w}(x,\Gn)}{\p t^2} =\frac{m^0_2-m^0_1}{m_1}g\,;\n
  x=an:~~~M_2 \f{\p^2  \bar{w}(\Gn)}{\p t^2} + \f{\Gvk_2}{\Gvk_1}\bar{w}(\Gn) -\f{1}{a}\gl[\bar{Q}^+(\Gn)-\bar{Q}^-(\Gn)\gr] =C\,,\n \bar{\CM}^+(\Gn)-\bar{\CM}^-(\Gn)=0\,,\n
C=g(1-\Gk_2/\Gk_1 + M_2 -M_1)\,. \eeq{tbabexe41}
In the following text, the  symbol ``bar'' are omitted for convenience of notation.
We also restrict our attention to the case where the damage is concentrated
at the  pillars, positioned at $x=na,$ and $m^0_1=m^0_2=m^0$.

\subsection{Interaction of the neighboring supported cross-sections}
We now consider the interaction of the neighboring cross-sections placed at the elastic support points, say, at $x=0$ and at $x=a$. For the steady-state regime the dynamic equation for the displacement of the beam placed between these cross-sections follows from \eq{tbabexe4} as
 \beq \f{\p^4 w(x,\Gn)}{\p x^4} + v^2\f{m^0}{m_1}  \f{\p^2 w(x,\Gn)}{\p \Gn^2} =0\,.\eeq{tbabexe5}
In terms of the Fourier transform on $\Gn$ with parameter $k$ we have
 \beq  \f{\p^4 w^F(x,k)}{\p x^4} -  v^2 k^2 \f{m^0}{m_1}w^F(x,k) = 0~~~\mbox{with }
v^2k^2=-\lim_{\Ge\to +0}(\Ge+\I v k)^2\,.
 \eeq{tbabexe6}
In the following, we will make use of an auxiliary solution of this equation corresponding to the boundary conditions (see \fig{Fig09}, panel II)
 \beq  w^F(0,k)= \gl(w^F(0,k)\gr)'=0\,,~~~w^F(a,k)=w_0\,,~~~ \gl(w^F(a,k)\gr)'=\Gf_0\,.\eeq{tbabexe7}
It is
 \beq w^F=W(x)w_0 +\GF(x)\Gf_0\,,\n
 W(x)=\f{(\cosh \Gl a-\cos \Gl a)(\cosh \Gl x-\cos \Gl x) - (\sinh \Gl a +\sin \Gl a) (\sinh \Gl x -\sin \Gl x)}{2(1-\cosh \Gl a\cos \Gl a)}\,,\n
 \GF(x)=\f{(\cosh \Gl a-\cos \Gl a)(\sinh \Gl x-\sin \Gl x) - (\sinh \Gl a -\sin \Gl a) (\cosh \Gl x -\cos \Gl x)}{2\Gl(1-\cosh \Gl a \cos \Gl a)}\,,\eeq{tbabexe8}
where $\Gl =(m^0/m_1)^{1/4} \sqrt{k v-\I 0}$. The transverse shear force, $Q=-w'''$, and the bending moment, $\CM=w''$, follow from this as
 \beq Q^F(0, \Gn) =Q_{w0}w_0 +Q_{\Gf 0}\Gf_0\,,~~~Q^F(a, \Gn) =Q_{wa}w_0 +Q_{\Gf a}\Gf_0\,,\n
  \CM^F(0,\Gn)= \CM_{w0}w_0 +\CM_{\Gf 0}\Gf_0\,,~~~\CM^F(a, \Gn) =\CM_{wa}w_0 +\CM_{\Gf a}\Gf_0\,,
  \eeq{tbabexe81a}
 with
 \beq
  Q_{w0} =  \f{\Gl^3(\sinh \Gl a +\sin \Gl a)}{1-\cosh \Gl a \cos \Gl a}\,,~~~Q_{\Gf 0}= -\f{\Gl^2(\cosh \Gl a-\cos \Gl a)}{1-\cosh \Gl a \cos \Gl a}\,,\n
  Q_{wa}=\f{\Gl^3(\cosh \Gl a \sin \Gl a+\sinh \Gl a \cos \Gl a)}{1-\cosh \Gl a \cos \Gl a}\,,~~~Q_{\Gf a}=-\f{\Gl^2\sinh \Gl a \sin \Gl a}{1-\cosh \Gl a \cos \Gl a}\,,\n
  \CM_{w0}=\f{\Gl^2(\cosh \Gl a - \cos \Gl a)}{1-\cosh \Gl a \cos \Gl a}\,,~~~\CM_{\Gf 0}=-\f{\Gl(\sinh \Gl a -\sin \Gl a)}{1-\cosh \Gl a \cos \Gl a}\,,\n
 \CM_{wa}=-\f{\Gl^2 \sinh \Gl a \sin \Gl a}{1-\cosh \Gl a \cos \Gl a}\,,~~~\CM_{\Gf a}=\f{\Gl(\cosh \Gl a \sin \Gl a -\sinh \Gl a \cos \Gl a)}{1-\cosh \Gl a \cos \Gl a}\,.\eeq{tbabexe81}

Note that for low speeds, $v\to 0$, the compliance coefficients tend to their static values, namely
 \beq Q_{w0}\to Q_{wa}\to \f{12}{a^3}\,,~~~Q_{\Gf 0}\to Q_{\Gf a}\to - \f{6}{a^2}\,,\n \CM_{w0}\to -\CM_{wa}\to \f{6}{a^2}\,,~~~\CM_{\Gf a}\to - 2\CM_{\Gf 0}\to \f{4}{a}\,.~~~\eeq{svotcc1}

\subsection{The Wiener-Hopf equation}
We now rearrange Eqs. \eq{tbabexe4}, \eq{tbabexe41} making the Fourier transform and using expressions (\ref{tbabexe81a},\ref{tbabexe81}).
We denote
 \beq \{w_+(k), \Gf_+(k)\} = \int_0^\infty \{w(\Gn), \Gf(\Gn)\}\E^{\I k \Gn}\D\Gn\,,\n \{w_-(k), \Gf_-(k)\} = \int_{-\infty}^0 \{w(\Gn), \Gf(\Gn)\}\E^{\I k \Gn}\D\Gn\,,\eeq{osFt1}
and deduce
 \beq (1-M_1v^2k^2)w_+(k) +(\Gk_2/\Gvk_1-M_2v^2k^2)w_-(k) +\f{2}{a}(Q_{wa}-Q_{w0}\cos ka )w^F(k)\n +\f{2\I}{a} Q_{\Gf 0}\sin ka \,\Gf^F(k)=\f{C}{0+\I k}\,,\n
 \I \CM_{w0}\sin ka\, w^F(k) + (\CM_{\Gf a}-\CM_{\Gf 0}\cos ka )\Gf^F(k)=0\,.\eeq{tge1}
Excluding  $\Gf^F(k)$ we obtain
 \beq L_1(k)w_+(k)+L_2(k)w_-(k) = \f{C}{0+\I k}= \f{C(L_1(k)-L_2(k))}{(0+\I k)[1-\Gk_2/\Gvk_1+(M_1-M_2)(0+\I k v)^2]}\,,\n
 L_1(k)=1+M_1(0+\I k v)^2 +\f{2}{a}\gl[(Q_{wa}-Q_{w0}\cos ka )+\f{Q_{\Gf 0}\CM_{w0}\sin^2 ka}{\CM_{\Gf a}-\CM_{\Gf 0}\cos ka }\gr]\,,\n
 L_2(k)=\f{\Gvk_2}{\Gvk_1}+M_2(0+\I k v)^2 +\f{2}{a}\gl[(Q_{wa}-Q_{w0}\cos ka )+\f{Q_{\Gf 0}\CM_{w0}\sin^2 ka}{\CM_{\Gf a}-\CM_{\Gf 0}\cos ka }\gr]\,.\eeq{pWHe1}
Finally, the Wiener-Hopf type equation follows in the form
  \beq L_0(k)w_+(k)+w_-(k)= \f{C}{(0+\I k)[1-\Gk_2/\Gvk_1+(M_1-M_2)(0+\I k v)^2]}[L_0(k) -1]\,,\n L_0(k)=L_1(k)/L_2(k)\,.\eeq{WHE1}

\subsection{Factorization}
To facilitate this action we assume that a small dissipation exists in proportion to the strain rate, that is, we correct the expression for the bending moment to be
\beq \CM = \f{\p^2 w(x,\Gn)}{\p x^2} + \Ga\f{\p^3 w(x,\Gn)}{\p t \p x^2}\,,\eeq{ioasv1}
where $\Ga$ is a small time parameter of viscosity. The non-dimensional beam equation after the Fourier transform on $\Gn$ becomes
 \beq (1+\I k v \Ga) \f{\p^4 w^F(x,k)}{\p x^4} -  v^2 k^2 w^F(x,k) = 0\,,\eeq{ioasv2}
that results in the change of the parameter $\Gl$
 \beq \Gl~~~  \Longrightarrow ~~~\f{\Gl}{(1+\I k v \Ga)^{1/4}}=\left(\frac{m^0}{m_1}\right)^{1/4}\f{\sqrt{ k v-\I 0}}{(1+\I k v \Ga)^{1/4}}= \mbox{O}\gl(|k|^{1/4}\gr)~~~(v>0, k\to \infty)\,.\eeq{ioasv3a}
Now the product
 \beq L(k)=\f{M_2}{M_1}L_0(k) \to 1 ~~~(k\to\pm\infty)\eeq{ioasv3}
has no zeroes
on the real $k$-axis (if $\Gk^0_2>0$). It means that its index is $0$, i.e.
\begin{equation}
\mbox{Ind} \,L(k)=\frac{1}{2\pi}[\mbox{Arg}L(\infty)-\mbox{Arg}L(-\infty)]=0\,.
\end{equation}
Besides, the integral of $\ln L(k)$ over the real $k$-axis converges.  This allows us to use the Cauchy type integral for the factorization of this function, that is to represent
 \beq L(k) = \lim_{\Im k\to 0}L_+(k)L_-(k)\,,~~~L_\pm(k)=\exp\gl[\pm\f{1}{2\pi\I}\inti \f{\ln L(\xi)}{\xi -k}\D\xi\gr]~~~ (\pm \Im k >0)\,.\eeq{fuCti1}
Eq. \eq{WHE1} can be represented in the form
 \beq L_+(k)w_+(k)+\f{M_2 w_-(k)}{M_1 L_-(k)}=G(k)\gl[\frac{M_1}{M_2}L_+(k) -\f{1}{L_-(k)}\gr]\,,\n
 G(k)= \f{M_2C}{M_1(0+\I k)[1-\Gk_2/\Gvk_1+(M_1-M_2)(0+\I k v)^2]}\,.\eeq{WHE1f1}

First consider the case $M_1=M_2=M$. The latter function can be split by two terms
 \beq G(k)\gl[L_+(k)-\f{1}{L_-(k)}\gr]=C_1+C_2\,,~~~C_1 = \f{g}{\I k}[L_+(k)-L_+(0)]\,,\n C_2 = \f{g}{0+\I k}\gl[L_+(0)-\f{1}{L_-(k)}\gr]\,,\eeq{WHE1f2}
where $C_{1,2}$ are regular in the upper and lower half-plane of $k$, \res.  The solution follows as
 \beq w_+(k) =\f{g}{\I k}\,\f{L_+(k)-L_+(0)}{ L_+(k)}\,,\n
 w_-(k)=\f{g}{0+\I k}[L_+(0)L_-(k)-1]\,.\eeq{WHE1f3}

In particular, using the limiting relations
 \beq w(\pm 0)=\lim_{k\to \pm\I\infty}(\mp\I k)w_\pm (k)\,,~~~L_\pm(\pm\I\infty)=1 \eeq{WHE1f4}
we find that
\beq w(+0)=w(-0)=w(0)= g[L_+(0)-1] \eeq{WHE1f5}
with
\beq L_\pm(0)=\sqrt{\f{\Gk_1}{ \Gk_2}}\exp\gl[\pm \f{1}{\pi}\int_0^\infty \f{\mbox{Arg}L(k)}{k}\D k\gr]~~~\gl(L(0)=\f{ \Gk_1}{ \Gk_2}\gr)\,.  \eeq{WHE1f6}

In the case $M_2<M_1$, the function $G(k)$ in \eq{WHE1f1} can be split by the corresponding terms as follows
 \beq G(k)=C_1(k)+C_2(k)\,,\n C_1(k)=C_0\gl[\f{L_+(k)-L_+(0)}{0+\I k}-\f{1}{2}\gl(\f{L_+(k)-L_+(\Gb)}{0+\I( k-\Gb)}+ \f{L_+(k)-L_+(-\Gb)}{0+\I( k+\Gb)}\gr)\gr]\,,\n
 C_2(k)=C_0 \gl[\f{L_+(0)}{0+\I k}-\f{1}{2}\gl(\f{L_+(\Gb)}{0+\I( k-\Gb)}+ \f{L_+(-\Gb)}{0+\I( k+\Gb)}\gr)\gr]-\f{G(k)}{L_-(k)}\,,\n
C_0= \f{M_2C}{M_1(1-\Gk_2/\Gvk_1)}\,,~~~\Gb=\f{1}{v}\sqrt{\f{1-\Gk_2/\Gvk_1}{M_1-M_2}}\,,\eeq{WHE1f6s}
and the solution is obtained as
 \beq w_+(k)= C_0 \gl[\f{1-L_+(0)/L_+(k)}{0+\I k}-\f{1}{2}\gl(\f{1-L_+(\Gb)/L_+(k)}{0+\I( k-\Gb)}+ \f{1-L_+(-\Gb)/L_+(k)}{0+\I( k+\Gb)}\gr)\gr]\,,\n
 w_-(k) = C_0 L_-(k)\gl[\f{L_+(0)}{0+\I k}-\f{1}{2}\gl(\f{L_+(\Gb)}{0+\I( k-\Gb)}+ \f{L_+(-\Gb)}{0+\I( k+\Gb)}\gr)\gr]- G(k)\,.\eeq{WHE1f6ss}

In particular, it follows that
\beq w(+0)=w(-0)=w(0)= C_0\gl[L_+(0)-\f{1}{2}\gl(L_+(\Gb)+L_+(-\Gb)\gr)\gr]\,.\eeq{WHE1f7}

We stress that the solution corresponds to the limit $\alpha\rightarrow 0$.

\section{The critical displacement $w(0)$}
We present in this section the dependance of the critical displacement $w(0)$ on the velocity of failure propagation $v$. This gives precise information on the range of velocity where steady-state regime is attainable or not, and where steady-state propagation is stable or unstable, information that cannot be obtained from the uniform continuous model of \az{Sect2}. Also, it is possible to establish a connection with subsonic, intersonic and supersonic regimes of the uniform continuous model considered in \az{Sect2} and find a universal character of the propagation which does not depend on some geometric parameters as the pillar normalized distance $a$.
In this work we focus our attention on numerical results where $m^0_1=m^0_2$, $M_1=M_2$ and $\Gvk_2/\Gvk_1<1$; the analysis of the model, where a contrast between the inertial properties of the damaged and undamaged structure is present, is left for a future work.

In the limit $\alpha=+0$, $\mbox{Arg}L(k)$ is a piecewise constant function over a semi-infinite
support ($v>0$), which makes the computation of the integral on an unbounded domain in Eq. \eq{WHE1f6} particularly difficult. To overcome this problem we consider a simplified model with massless beams where the inertial effects are concentrated in correspondence of the elastic supports (shown in \fig{Fig01}b) and a second one where we introduce a small positive dissipative term $\alpha$ (see Eq. \eq{ioasv1}).

\subsection{The lattice beam model for masses concentrated at nodal points}
We evaluate the displacement according to formula \eq{WHE1f5}, which also include evaluation of the $\mbox{Arg}[L(k)]$ and of the integral in Eq. \eq{WHE1f6}. This results will enable us to judge on the values of the speed of the propagating fault in the discrete flexural system. We consider the case when the normalised masses $M_1$ and $M_2$ used in the previous section are now replaced by the total redistributed mass, which includes the added masses assigned to junction points $x=na$ and the mass assigned earlier to the beam between the neighbouring supports, as (in the non-dimensional form)
$M_1+m^0/m_1=1$ and $M_2+m^0/m_1 =1-M_1+M_2$, respectively. In this section, we assume that $M_1$ and $M_2$ are equal, and we use the notation $M = M_1 = M_2$.
As a result, we deal with a system of unit point masses placed  at $x=an$ and connected by non-inertial elastic beams.
In this case the equation \eq{tbabexe5} reads as
\beq
\f{\p^4 w(x,\Gn)}{\p x^4}  =0\,.
\eeq{dsp_001}
and hence, when $an<x<a(n-1)$, the displacement is cubic in $x$. For $x=an$ the equations are the same as in \eq{tbabexe4} and \eq{tbabexe41}.

\begin{figure}[!ht]

\centering
\vspace*{10mm} \rotatebox{0}
{\resizebox{!}{9.cm}{\includegraphics[scale=0.5]{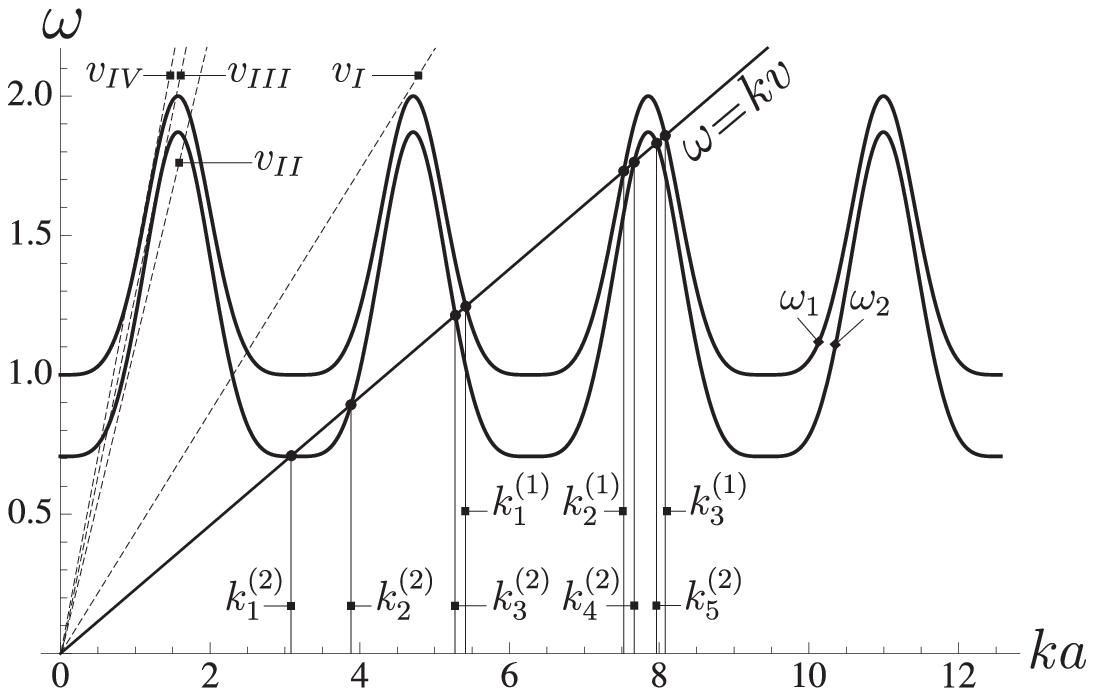}}}
\caption{
Dispersion diagrams $\omega_1(ka)$, $\Gvk=\Gvk_1$, and $\omega_2(ka)$,  $\Gvk=\Gvk_2$,  for waves in the discrete flexural system, Fig. \ref{Fig01}b.  The diagrams are given for stifness ratio $\Gvk_2/\Gvk_1=0.5$ and normalized span length $a=2.0$. The straight line $\omega=k\,v$ ($v=0.23$) has a finite number of intersections (denoted with small black circles) at wavenumbers $k^{(1)}_1,\ldots,k^{(1)}_3$ and $k^{(2)}_1,\ldots,k^{(2)}_5$  with the two dispersion curves $\omega_1$ and $\omega_2$, \res. The velocities $v_{I}$, $v_{II}$, $v_{III}$, $v_{IV}$ are reported in the critical displacement diagram in \fig{Fig12}.}
    \label{Fig10}
\end{figure}

The Wiener-Hopf equation for the one-sided Fourier transform of the displacement  has the form \eq{WHE1} with $M_1=M_2$. Taking into account Eq. \eq{svotcc1} we deduce
\beq
L_1(k)=1+(0+ikv)^2+\frac{12}{a^4}\frac{(1-\cos ka)^2}{2+\cos ka},\n
L_2(k)=\Gvk_2/\Gvk_1+(0+ikv)^2+\frac{12}{a^4}\frac{(1-\cos ka)^2}{2+\cos ka}.
\eeq{dsp_002}
Consequently, the dispersion equation describing the waves for $\eta>0$ and $\eta<0$ are
\beq
\omega_1=\sqrt{1+\frac{12(1-\cos ka)^2}{a^4(2+\cos ka)}}\,,\n
\omega_2=\sqrt{\f{\Gvk_2}{\Gvk_1}+\frac{12(1-\cos ka)^2}{a^4(2+\cos ka)}}\,,
\eeq{dsp_003}
\res. The corresponding dispersion curves are shown in \fig{Fig10}.

\begin{figure}[!ht]

\centering
\vspace*{10mm} \rotatebox{0}
{\resizebox{!}{9.5cm}{\includegraphics[scale=0.5]{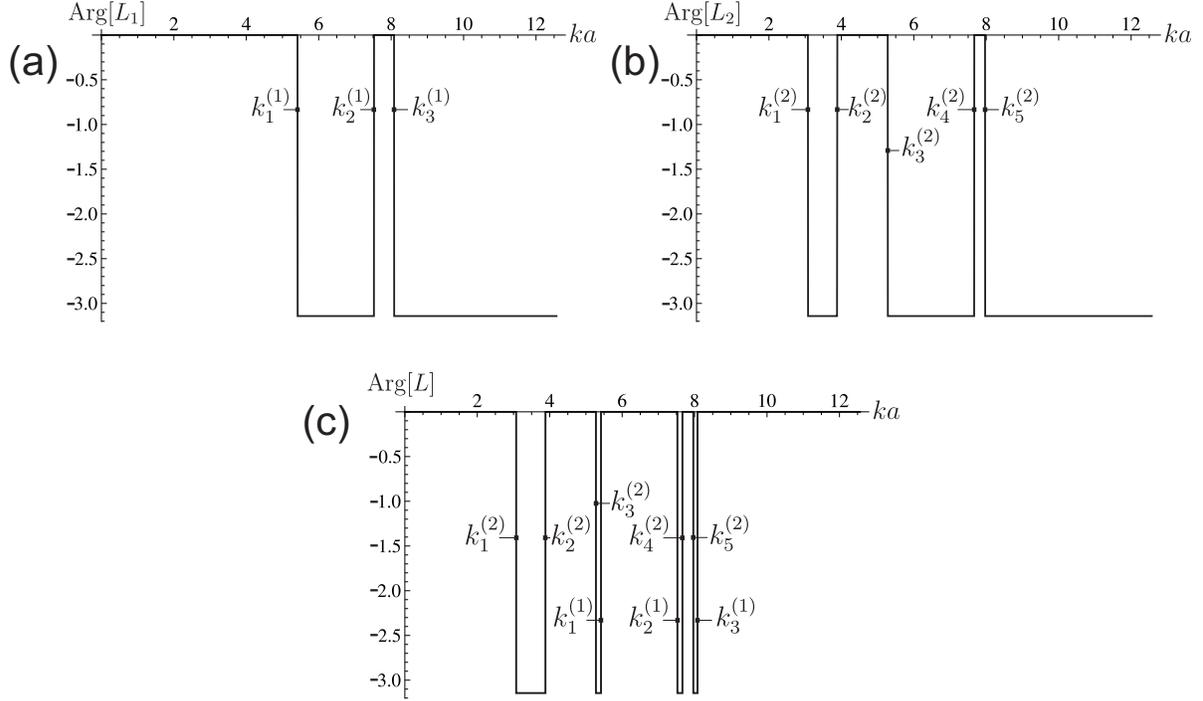}}}
\caption{The piecewise constant (a) $\mbox{Arg}[L_1]$, (b) $\mbox{Arg}[L_2]$ and (c) $\mbox{Arg}[L]$ as a function of the normalized wavenumber $ka$. Results are given for $a=2.0$ and $v=0.23$. Jumps occur at the intersection point wavenumbers $k^{(1)}_1,\ldots,k^{(1)}_3$ and $k^{(2)}_1,\ldots,k^{(2)}_5$ highlighted in \fig{Fig10}.}
    \label{Fig11}
\end{figure}

Due to Eq. \eq{ioasv3} $L_0(k)=L(k)=L_1(k)/L_2(k)$. In this case, $\mbox{Arg}[L(k)]$ is a piecewise constant function with a finite, speed-dependent support ($v>0$) vanishing in the neighborhood of the origin.
The jumps occur at the points of intersection at $k=k^{(1)}_j$, $k=k^{(2)}_j$ between the ray $\Go = kv$ and the dispersion curves $\Go_{1,2}(k)$, as shown in \fig{Fig10}. The graphs of $\mbox{Arg}[L_1(k)]$, $\mbox{Arg}[L_2(k)]$ and $\mbox{Arg}[L(k)]$ for $v=0.23$ are given in \fig{Fig11}. According to Eqs. \eq{WHE1f5} and \eq{WHE1f6} the displacement at the origin is given by the formula
\beq
w(0)=g\left[\sqrt{\frac{\Gvk_1}{\Gvk_2}} \left(\prod_{i=1}^{n_1} \frac{k^{(1)}_{2i}}{k^{(1)}_{2i-1}}\prod_{j=1}^{n_2}\frac{k^{(2)}_{2j-1}}{k^{(2)}_{2j}}\right)
\frac{k^{(2)}_{2n_2+1}}{k^{(1)}_{2n_1+1}}-1\right],
\eeq{dsp_004}
where $2n_1+1$ and $2n_2+1$ are the number of intersections of the straight line $\omega=kv$, with the two dispersion curves $\omega=\omega_1(k)$ and $\omega=\omega_2(k)$, respectively.
Note that in \fig{Fig10} and \fig{Fig11}, at $v=0.23$, $n_1=1$ and $n_2=2$.

\begin{figure}[!ht]

\centering
\vspace*{10mm} \rotatebox{0}
{\resizebox{!}{9.cm}{\includegraphics[scale=0.5]{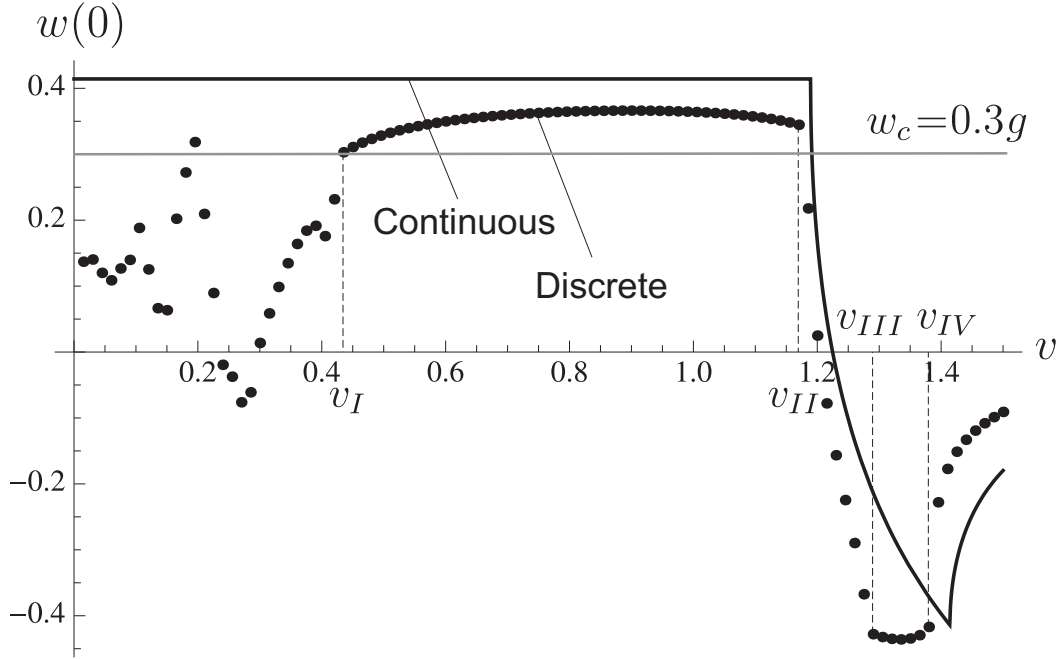}}}
 \caption{Critical displacement $w(0)$ as a function of the speed $v$. The uniform continuous model (\fig{Fig01}a, eqs. \ref{bee10}, \ref{bee13} and \ref{bee14w}) is compared with the discrete massless beam model (\fig{Fig01}c, eq. \ref{dsp_004}). Results are given for $\Gvk_2/\Gvk_1=0.5$ and in the discrete model $a=2$. The velocities $v_{I}$, $v_{II}$, $v_{III}$, $v_{IV}$ are reported in the dispersion diagram in \fig{Fig10}. The critical displacement $w_c/g=0.3$ is denoted by the grey line.}
    \label{Fig12}
\end{figure}

The critical displacement $w(0)$ as a function of the speed $v$ is given in \fig{Fig12} for the stiffness ratio $\Gvk_2/\Gvk_1=0.5$ and span length $a=2$ and compared with the continuous beam model on elastic foundation. It is evident the jagged structure for $v<v_I$, where the straight line $\omega=kv$ has multiple intersection with the dispersion curves $\omega_1$ and/or $\omega_2$ and, in principle, different waves can radiate from $\eta=0$. At $v=0.23$ waves can radiate with wavelength $k=k^{(1)}_2$ in the region $\eta>0$ and with wavelengths $k=k^{(2)}_1,k^{(2)}_3,k^{(2)}_5$ in the region $\eta<0$ (see \fig{Fig10}).
In this low-speed region more and more sinusoidal waves arise as the speed decreases, with intensities extremely sensitive to the speed, causing irregularities in the dependence of the critical displacement on $v$.

However, additional conditions have to be fulfilled by displacement \eq{dsp_004} to be a steady-state solution of the problem \eq{tbabexe4}-\eq{tbabexe41}. As noted by Marder and Gross (1995) for the problem of dynamic crack propagation in lattices, the solution is linearly unstable whenever $w(0)$ is an increasing function of $v$, which means that the failure wave accelerates. Therefore, all the increasing portions of the curve can be ruled out.

But this is not enough, a further condition of admissibility needs to be satisfied by $w(0)$: it is not only necessary to verify that the displacement $w(0)$ has reached the critical value $w_c$ (positive in normalized coordinates) at $t=0$, but this must be the first time at which the displacement  reaches this critical value. This is equivalent to check that $w(\eta)<w_c$ for $\eta>0$. Instead of looking at the distribution of the displacement for $\eta>0$, Slepayn and Ayzenberg-Stepanenko (2004) considered a necessary condition equivalent to
\begin{equation}
\frac{dw(\eta)}{d\eta}=-\frac{1}{v}\frac{\partial w}{\partial t}<0\qquad(\eta=0),
\label{eq400}
\end{equation}
for the problem of transition waves in bistable-bond lattices.
Considering the limiting relation
\begin{equation}
\dot{w}(0) = \lim_{k\to\I\infty}(-\I k)(\I k v)\gl[w_+(k) - \f{w(0)}{(-\I k)}\gr]
\label{eq401}
\end{equation}
and that for $k\to +\I\infty$
\begin{equation}
\f{1}{L_+(k)} \sim 1 -  \f{1}{2\pi\I}\inti\f{\ln L(\xi)}{\xi-k} \D\xi\,,
\label{eq402}
\end {equation}
we check the range of velocity where
\begin{equation}
\dot{w}(0)=\f{v\, w(0)}{2\pi}\inti \ln L(k)\D k >0.
\label{eq403}
\end {equation}
By using the symmetries of $L_1(k)$ and $L_2(k)$, the integral in \eq{eq403} is given by
\begin{equation}
\inti \ln L(k)\D k=2 \int_0^\infty \ln (|L(k)|)\D k.
\label{eqA01}
\end{equation}
which is integrable but the integrand has logarithmic singularities. Then, the integral can be split as
\begin{equation}
\int_0^\infty = \int_0^{k_1}+\int_{k_1}^{k_2}+\ldots+\int_{k_{2(n_1+n_2+1)}}^{k_M}+\int_{k_M}^\infty
\label{eqA02}
\end{equation}
where $(k_1,k_2,\ldots,k_{2(n_1+n_2+1)})$ are the $2n_1+2n_2+2$ intersection points between the line $\omega=k\,v$ and the dispersion curves $\omega_1$ and $\omega_2$, such that $k_i<k_j$ for $i<j$. For the parameters considered in \fig{Fig10}, at $v=0.23$, the wavenumbers are $(k^{(2)}_1,k^{(2)}_2,k^{(2)}_3,k^{(1)}_1,k^{(1)}_2,k^{(2)}_4,k^{(2)}_5,k^{(1)}_3)$.
The first  $2n_1+2n_2+2$ integrals can be computed numerically using standard quadrature rules for integrand having logarithmic singularities at the endpoints and $k_M$ has been chosen to be sufficiently large to guarantee that $|\int_{k_M}^\infty \ln (|L(k)|)\D k|/|\int_0^{k_M} \ln (|L(k)|)\D k|<10^{-6}$.

These values are reported in Table \ref{Table01}.
We see that for $v\leq 0.55$, $\dot w(0)<0$ (except at $v=0.2$), showing that in this range no steady-state propagation, $w=w(x,\eta)$, can exist.

\begin{table}[h!]
  \begin{center}
  \begin{tabular}{|c|c|}
\hline
$v$ & $\dot w(0)$ \\
\hline
$0.10$	&	$-0.35*10^{-3}$	\\	
$0.15$	&	$-0.25*10^{-3}$	\\	
$0.20$	&	$0.12*10^{-1}$		\\	
$0.25$	&	$-0.12*10^{-2}$	\\	
$0.30$	&	$-0.80*10^{-3}$	\\	
$0.35$	&	$-0.88*10^{-2}$	\\
$0.40$	&	$-0.82*10^{-2}$	\\	
$0.45$	&	$-0.18*10^{-1}$	\\	
\hline
\end{tabular}
\hspace{4 mm}
  \begin{tabular}{|c|c|}
\hline
$v$ & $\dot w(0)$ \\
\hline
$0.50$	&	$-0.88*10^{-2}$	\\	
$0.55$	&	$-0.25*10^{-2}$	\\	
$0.60$	&	$0.31*10^{-2}$		\\	
$0.65$	&	$0.83*10^{-2}$		\\	
$0.70$	&	$0.13*10^{-1}$		\\	
$0.75$	&	$0.19*10^{-1}$		\\
$0.80$	&	$0.24*10^{-1}$		\\	
$0.85$	&	$0.30*10^{-1}$		\\	
\hline
\end{tabular}
\hspace{4 mm}
  \begin{tabular}{|c|c|}
\hline
$v$ & $\dot w(0)$ \\
\hline
$0.90$	&	$0.36*10^{-1}$		\\	
$0.95$	&	$0.43*10^{-1}$		\\	
$1.00$	&	$0.52*10^{-1}$		\\	
$1.05$	&	$0.62*10^{-1}$		\\	
$1.10$	&	$0.76*10^{-1}$		\\	
$1.15$	&	$0.10*10^{0}$		\\
$1.175$	&	$0.77*10^{-1}$		\\	
$1.2$	&	$0.11*10^{-1}$		\\	
\hline
\end{tabular}
\end{center}
\caption{Value of the displacement time derivative $\dot w(0)$ at different speeds $v$.}
\label{Table01}
\end{table}

\begin{figure}[!ht]

\centering
\vspace*{10mm} \rotatebox{0}
{\resizebox{!}{9.cm}{\includegraphics[scale=0.5]{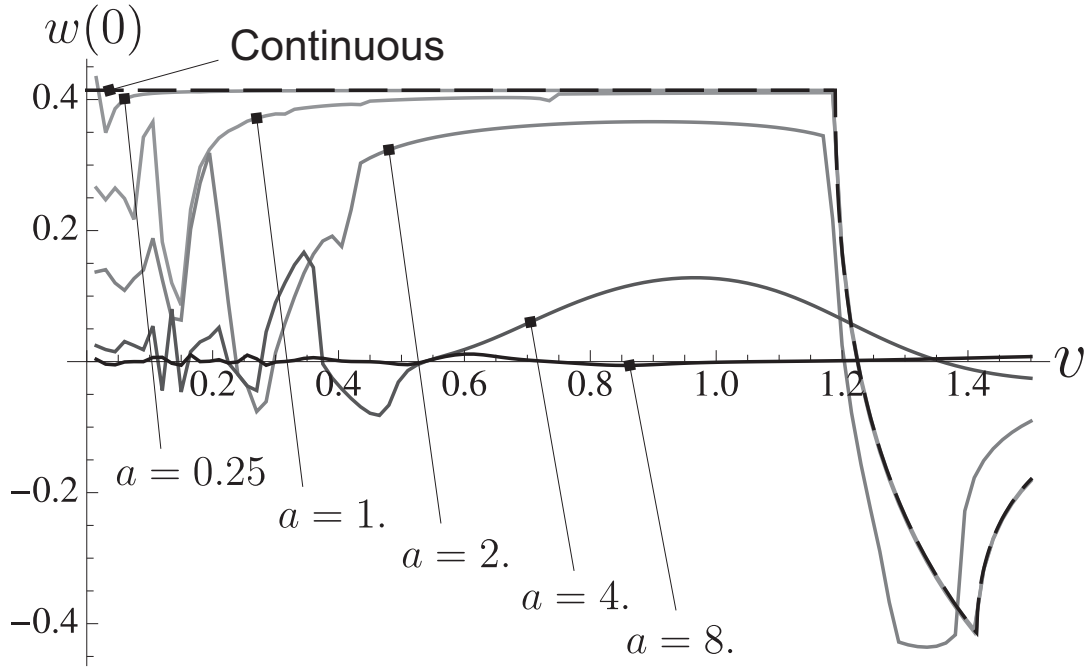}}}
\caption{Critical transversal displacement $w(0)$ as a function of the speed $v$. The discrete massless beam model (continuous grey lines) and the uniform continuous model (dashed black line) are represented with
$\Gvk_2/\Gvk_1=0.5$. In the discrete model $a=0.25,\,1.0,\,2.0,\,4.0,\,8.0$.}
    \label{Fig13}
\end{figure}

Following the behavior of the transversal displacement $w(0)$ at higher velocities, it can be seen that there is a sort of plateau in the range $v_{I}<v<v_{II}$ with a drastic drop in the range $v_{II}<v<v_{III}$.  This dependence shows strong similarities with the behavior of the continuous model.


We can conclude that \emph{steady-state propagation, when possible, occurs in a range of velocity which is exceptionally well approximated by the narrow range of velocities  \eq{v_constr}, within the intersonic regime, provided by the continuous model of beam on elastic foundation}. In such a velocity range, the steady state solution for the discrete case is linearly stable and satisfies the conditions \eq{eq403} and the critical one $w(0)=w_c$.

In \fig{Fig13} the displacement $w(0)$ is given as a function of velocity $v$ for different values of the normalised span length $a$. We note the good agreement between the continuous and the discrete models for sufficiently small values of $a$, which is expected on physical ground, an issue that will be better discussed in \az{Sect42}. Also,
the continuous model is approached from below at decreasing values of $a$.

\subsubsection{Dispersion diagram transformation and the continuous limit}

\begin{figure}[!ht]

\centering
\vspace*{10mm} \rotatebox{0}
{\resizebox{!}{8.cm}{\includegraphics[scale=0.5]{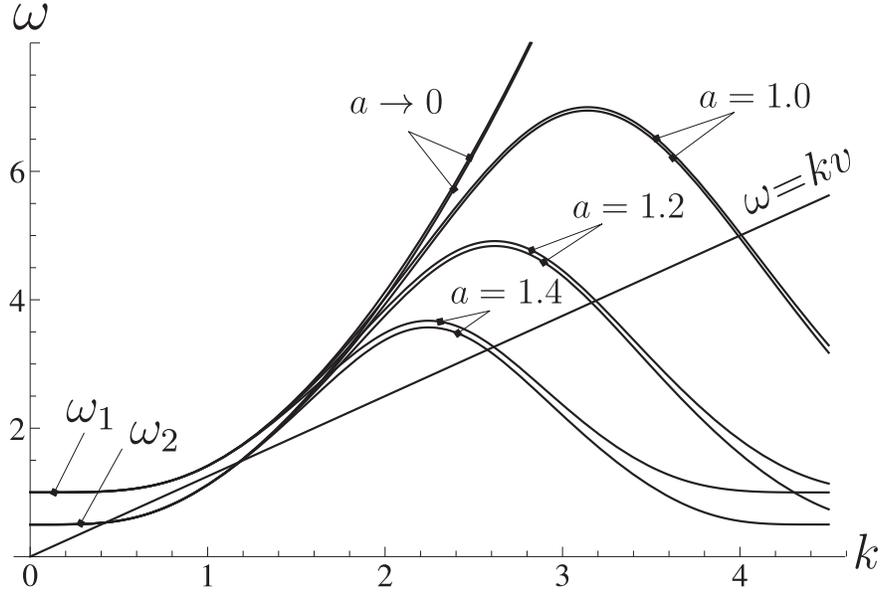}}}
 \caption{Dispersion diagrams $\omega_{1,2}(k)$ for the discrete massless beam model. Results are given for $a=1.4,\,1.2,\,1.0,\,+0$ and $\Gvk_2/\Gvk_1=0.5$.}
    \label{Fig14}
 \end{figure}

Let $a \to +0.$ Expanding in $ka$ we deduce from \eq{dsp_003} that in any finite range of the wavenumber the dispersion relations take the form
\beq
\omega_{1} \sim \sqrt{1+k^4}\,,~~~
\omega_{2} \sim \sqrt{\Gvk_2/\Gvk_1+k^4}\,,
\eeq{dsp_003a}
while the remaining periodic part of the dispersion curves moves away to infinity (see \fig{Fig14}). In this transformation, the two branches, $\Go_1$ and $\Go_2$   \eq{dsp_003}, coalesce, becoming closer and closer, and the separation $\Go_1(k) - \Go_2(k) \to 0$  as $a \to 0$. Hence in the limit the cross points other than those corresponding  to the asymptotic formulae \eq{dsp_003a}, do not contribute to the product in \eq{dsp_004} for the displacement $w(0)$.
At the same time, these asymptotic expressions coincide with the dispersion relations \eq{bee7} derived for the continuous model, which has the same average density and the same average support stiffness as the discrete model.  Thus, the dynamic response of the discrete flexural system is shown to converge to the continuum limit as $a \to 0.$

\subsection{Inertial beam model}
\label{Sect42}

For the case of dynamic beam on the set of discrete elastic supports (\fig{Fig01}c) the dispersion diagrams for the undamaged and damaged structures have infinite numbers of curves (see, for example, Brun et al., 2012) and Eq. \eq{dsp_004} is still valid, but with $n_1,n_2\rightarrow \infty$.  To compute the displacement $w(0)$ is then convenient to introduce some dissipation as in \eq{ioasv1} and make use of Eqs. \eq{WHE1f5} and \eq{WHE1f6}.

In \fig{Fig15} the displacement $w(0)$ is given as a function of velocity $v$ for different values of the normalised span length $a$. Computations have been done assuming
$\alpha=0.01$ and the integral in \eq{WHE1f6} was truncated imposing a relative error smaller than $10^{-6}$.  The comparative analysis between the results of \fig{Fig13} and \fig{Fig15} shows larger differences at increasing values of $a$, where the difference between the coefficients in \eq{tbabexe81} and in \eq{svotcc1} is more pronounced and the wave propagation is better described by the beam inertial model.

\begin{figure}[!ht]

\centering
\vspace*{10mm} \rotatebox{0}
{\resizebox{!}{9.cm}{\includegraphics[scale=0.5]{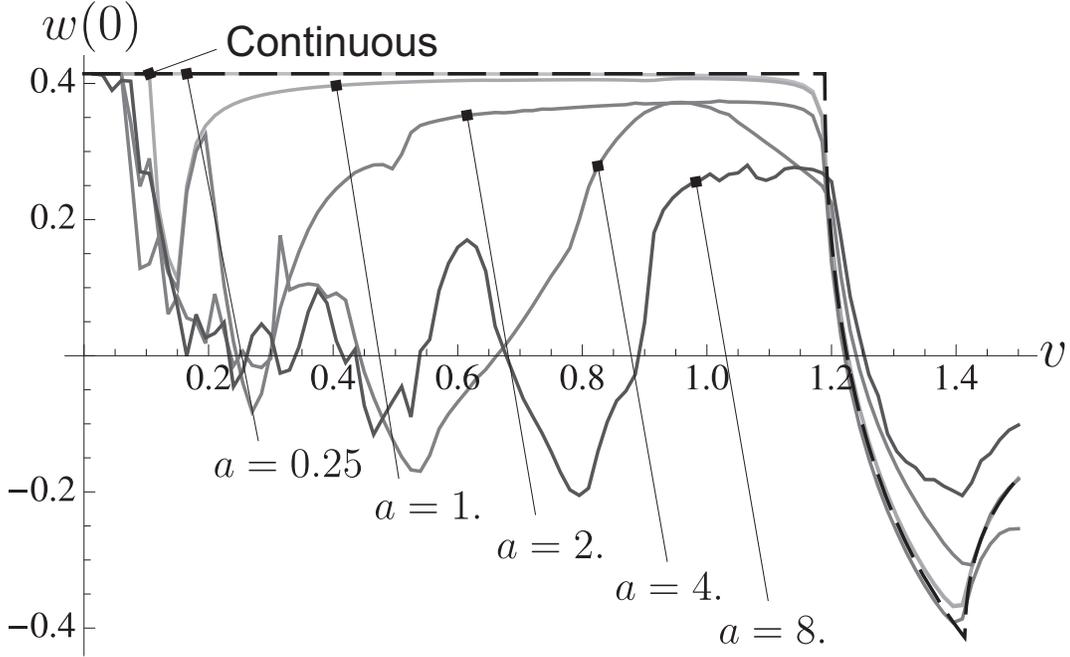}}}
\caption{Critical displacement $w(0)$ as a function of the speed $v$. The discrete inertial beam model (continuous grey  lines) and the uniform continuous model (dashed black line) are represented with
$\Gvk_2/\Gvk_1=0.5$. In the discrete-continuous model $m^0/m_1=0.67$ and $a=0.25,\,1.0,\,2.0,\,4.0,\,8.0$.}
    \label{Fig15}
\end{figure}

Despite of the fact that the interval of velocity where oscillations are present increases with $a$, all the curves display a drastic drop in correspondence of the intersonic regime identified by the model of beam on elastic foundation, thus giving a \emph{universal property for the steady-state failure propagation} in such mono-dimensional structures.

Finally, we report some data deduced from real life bridges, showing the range of possible engineering applications of the proposed model. We first consider a bridge of average dimension, the S'Adde bridge, whose geometrical and material properties are given in the caption of Fig. 4 of Brun et al., 2012, with a span length of $90$ m. In such a case the normalised span length is $a=2.02$ for vertical flexural waves and $a=0.83$ for horizontal flexural waves. In addition, we also pay attention to the Millau viaduct, one of the tallest vehicular bridge and the longest multiple cable-stayed bridge in the world (Magalh$\tilde{\mbox{a}}$es et al., 2012). The steel structure has Young Modulus $E=210000$ MPa, moments of inertia $J_y=1137.7$ m$^4$ and $J_z=8267.5$ m$^4$, vertical and transverse stiffness $\Gvk_z = 20000$ MPa m and $\Gvk_y =41.84$ MPa m, respectively, and span length of $342$ m. Then, the structure has normalised span length $a=7.61$ for vertical flexural waves and $a=
 0.99$ for the horizontal ones, showing that, also for this extreme structure, we are in the range of $a$  where the presented model can be used to model the steady-state failure propagation.

\section{Concluding remarks}

We have obtained novel results on an advance of a transition flexural wave through a beam-like periodically supported discrete structure. Both propagating and evanescent waves are included in a general solution and hence influence the interaction between different nodal points within the discrete system.

It was shown that in the considered problem, in contrast to the `classical' fracture in continuum mechanics, the energy release under the support damage remains positive not only in the case where there is no wave radiation but also in a part of the intersonic speed range. In this case, a part of the total energy released from the heavy beam is radiated by an elastic wave excited by the jump in the stiffness of the support.

The model has several significant applications in structural mechanics, including safety evaluation and design algorithms for earthquake resistant long bridges.
The necessary condition, reduced to evaluation of the critical displacement,  under which a failure of the supporting piles of a bridge  can propagate,
has been identified and analysed for a range of physical parameters. The failure has been modeled as a drop of the stiffness or of the total mass the support. For the model of inertial continuous beam on distributed elastic foundation the solution has been obtained by solving directly the corresponding equations of motion for steady-state propagation and checking, in addition, the energy balance. Computational examples, as outlined in the above section, strongly suggest the applicability of the proposed model to real life bridge systems such as the S'Adde bridge and the Millau viaduct.

We note that the model discussed in the paper is radically different from models describing crack propagation in discrete lattices, both in scalar (anti-plane shear) and vector cases. The present models deal with the fourth-order flexural wave equation, whose dispersion properties include unique features, which are absent in the two-dimensional lattice crack models.

A regime, which we refer to as intersonic, has been identified ($\sqrt{2}{(\Gvk)}^{1/4}<v<\sqrt{2}$), and it has been shown that the solution corresponding to the steady propagation of the interface wave is intersonic.
An efficient and elegant mathematical approach based on analysis of a functional equation of the Wiener-Hopf type, has led to the expression of the displacement on the interface wave to be written as a product of terms evaluated directly from analysis of the kernel function of the Wiener-Hopf equation. The discrete model has been compared with the simpler continuous approach, and of course it has been shown that the lattice solution possesses new feature, not reported in the past for interfacial flexural waves.

Lastly, the conditions allowing the failure wave to propagate and the characteristic wave speeds are found.

\vspace{5mm}
\noindent {\bf Acknowledgements.} This paper was partially written when L.S.  and A.B.M.  were Visiting Professors at Cagliari University under the 2011 program funded by Regione Autonoma della Sardegna. L.S. is also thankful for the support provided by FP7-People-2011-IAPP EU Grant No. 284544. M.B. acknowledges the support of
of the EU FP7 Grant  No. PIEF-GA-2011-302357. We particularly thank Dr. Eng. F. Giaccu for the estimation of the geometric parameters of the Millau viaduct.

\newpage
\vskip 18pt
\begin{center}
{\bf  References}
\end{center}
\vskip 3pt

\inh Balk, A.M., Cherkaev, A.V., and  Slepyan, L.I., 2001a.
Dynamics of Chains with Non-monotone Stress-Strain Relations. I. Model and Numerical Experiments.
J. Mech. Phys. Solids 49, 131-148.

\inh Ba$\check{z}$ant, Z.P., and Zhou, Y., 2002.
Why Did the World Trade Center Collapse? Simple Analysis.
J. Engn. Mech., 2-6.

\inh Ba$\check{z}$ant, Z.P., and Verdure, M., 2007.
Mechanics of Progressive Collapse: Learning from World Trade Center and Building Demolitions.
J. Engn. Mech., 308-319.

\inh Ba$\check{z}$ant, Z.P., ASCE, H.M., Le, J-L., Greening, F.R., and Benson, D.B., 2008.
What Did and Did Not Cause Collapse of World Trade Center Twin Towers in New York?
J. Engn. Mech., 892-906.

\inh Balk, A.M., Cherkaev, A.V., and  Slepyan, L.I., 2001b.
Dynamics of Chains with Non-monotone Stress-Strain Relations. II. Nonlinear Waves and Waves of Phase Transition.
J. Mech. Phys. Solids 49, 149-171.

\inh Brun, M., Giaccu, G.F.,  Movchan, A. B., and Movchan, N. V., 2012.
Asymptotics of eigenfrequencies in the dynamic response of elongated multi-structures.
Proc. R. Soc. London A  468,  378-394.

\inh Buehler, M.J., Abraham, F.F., and Gao H., 2003,
Hyperelasticity governs dynamic fracture at a critical length scale.
Nature 426, 141-146.

\inh Cherkaev, A., Cherkaev, E. and Slepyan, L, 2005. Transition
waves in bistable structures. I. Delocalization of damage.
J. Mech. Phys. Solids 53(2), 383-405.

\inh Galin L.A. and Cherepanov G.P. (1966)
Self-sustaining Failure of a Stressed Brittle Body.
Sov. Phys. Dokl. 11(3): 367-369.

\inh Grigoryan, S.S. (1967)
Some Problems of the Mathematical Theory of Deformation and Fracture of Hard Rocks.
Appl. Math. Mech. (PMM) 31(4): 667-686.

\inh Marder, M., Gross, S. (1965)
Origin of crack tip instabilities.
J. Mech. Phys. Solids 43(1): 1-48.

\inh Magalh$\tilde{\mbox{a}}$es, F., Caetano, E., Cunha, \'A., Flamand, O., Grillaud, G., (2012)
Ambient and free vibration tests of the Millau Viaduct: Evaluation of alternative processing strategies.
Eng. Struct., 45: 372-384.

\inh Ngan S.-C. and Truskinovsky L. (1999)
Thermal Trapping and Kinetics of Martensitic Phase Boundaries.
J. Mech. Phys. Solids 47: 141-172.

\inh Puglisi G. and Truskinovsky L. (2000)
Mechanics of a Discrete Chain with Bi-Stable Elements.
J. Mech. Phys. Solids 48: 1-27.

\inh Slepyan L.I. (1968)
Brittle Failure Waves.
Mech. Solids 3(4): 202-204.

\inh Slepyan L.I. (1977)
Models in the Theory of Brittle Fracture Waves.
Mech. Solids 12(1): 170-175.

\inh Slepyan L.I. and Troyankina L.V. (1969)
The Failure of Brittle Bar Under Longitudinal Impact.
Mech. Solids 4(2): 57-64.

\inh Slepyan L.I. and Troyankina L.V. (1984)
Fracture Wave in a Chain Structure.
J. Appl. Mech. Techn. Phys. 25(6): 921-927.

\inh Slepyan L.I. and Troyankina L.V. (1988)
Shock Waves in a Nonlinear Chain.
In: Gol'dstein R.V. (Ed) Plasticity and Fracture of Solids. Nauka, Moscow, 175-186 (in Russian).

\inh Slepyan L.I. (2000)
Dynamic Factor in Impact, Phase Transition and Fracture.
J. Mech. Phys. Solids 48: 931-964.

\inh Slepyan L.I. (2001)
Feeding and Dissipative Waves in Fracture and Phase Transition. II. Phase-transition Waves.
J. Mech. Phys. Solids 49: 513-550.

\inh Slepyan, L.I., 2002.
Models and Phenomena in Fracture
Mechanics. Springer, Berlin.

\inh Slepyan, L.I., and Ayzenberg-Stepanenko, M.V., 2004.
Localized transition waves in bistable-bond lattices.
J. Mech. Phys. Solids 52(7), 1447-1479.

\inh Slepyan, L., Cherkaev, A. and Cherkaev, E, 2005.
Transition waves in bistable structures. II. Analytical solution: wave speed and energy dissipation.
J. Mech. Phys. Solids 53(2), 407-436.

\inh Truskinovsky L. (1994)
About the ``Normal Growth" Approximation in the Dynamic Theory of Phase Transitions.
Cont. Mech. Thermodyn 6: 185-208.

\inh Truskinovsky L. (1997)
Nucleation and Growth in Elasticity.
In: Duxbury P. and Pence T. (Eds) Dynamics of Crystal Surfaces and Interfaces, Plenum Press, New York, 185-197.

\inh  Vainchtein, A. and Kevrekidis, P.G., 2012. Dynamics of phase transitions in a piecewise linear
diatomic chain. J Nonlinear Sci. 22, 107-134.

\end{document}